\documentclass[11pt,a4paper]{article}
\usepackage[T1]{fontenc}
\usepackage{xcolor}

\usepackage{eurosym}
\usepackage{makecell} 
\usepackage{threeparttable}
\usepackage{amssymb}

\usepackage[a4paper,top=2cm,bottom=2cm,left=2cm,right=2cm,marginparwidth=1.75cm]{geometry}
\usepackage{tikz}
\usepackage{booktabs}
\usepackage{tabularx}     
\usepackage{amsmath}
\usepackage{graphicx}
\usepackage{hyperref}
\usepackage{tikz}
\usetikzlibrary{arrows.meta, positioning, calc}

\tikzset{
  box/.style={rectangle, draw, rounded corners, align=center, minimum width=3.8cm, minimum height=1cm},
  arrow/.style={-{Latex}, thick}
}

\usepackage[affil-it]{authblk}

\usepackage{apacite}
\usepackage{natbib}

\usepackage{setspace}    
\usepackage{lineno}      

\title{\bfseries \LARGE A Modeling and Optimization Framework for Fostering Modal Shift through the Integration of Tradable Credits and Demand-Responsive Autonomous Shuttles}
\author[1]{Zenghao Hou*}
\author[2]{Ludovic Leclercq}
\affil[1]{Postdoctoral researcher, Université Gustave Eiffel, ENTPE, EMob-Lab, Lyon, France}
\affil[2]{Professor, Université Gustave Eiffel, ENTPE, EMob-Lab, Lyon, France}
\date{\vspace{-5ex}}

\begin{document}
\maketitle
\hrule\vspace{0.5em}
\section*{ABSTRACT}
Tradable Credit Schemes (TCS) promote the use of public and shared transport by capping private car usage while maintaining fair welfare outcomes by allowing credit trading. However, most existing studies assume unlimited public transit capacity or a fixed occupancy of shared modes, often neglecting waiting time and oversimplifying time-based costs by depending solely on in-vehicle travel time. These assumptions can overstate the system's performance with TCS regulation, especially when there are insufficient public or shared transport supplies.

To address this, we develop a dynamic multimodal equilibrium model to capture operation constraints and induced waiting times under TCS regulation. The model integrates travelers' mode choices, credit trading, traffic dynamics, and waiting time, which depend on key operational features of service vehicles such as fleet size and capacity. 

Besides, most TCS studies assume fixed transport supply, overlooking supply-side responses triggered by demand shifts. Therefore, we further propose integrating adaptive supply management through the deployment of Demand-Responsive Autonomous Shuttles (DRAS) and develop a bi-level optimization framework that incorporates the equilibrium model to jointly optimize TCS design and operational strategies for the DRAS.

We apply the framework to a section of the A10 highway near Paris, France, to examine demand–supply interactions and assess the potential benefits of jointly implementing TCS and DRAS. Numerical results demonstrate the importance of modeling operational features within multimodal equilibrium and incorporating flexible supply in TCS policies for mitigating overall generalized cost.

\textbf{Keywords}: Multimodal Stochastic User Equilibrium, Multimodal Macroscopic Fundamental Diagram, Bi-level Optimization with Equilibrium, Demand-Responsive Autonomous Shuttle, Tradable Credit Scheme, Public Transportation.
\vspace{1.5em}\hrule
\section{Introduction}
The growing mismatch between vehicular travel demand and limited road capacity has led to severe congestion, calling for innovative solutions to balance demand and capacity. Cities, such as New York \citep{cook2025short} and London \citep{tang2021cost}, adopted congestion pricing strategies to raise the monetary cost of driving vehicles in downtown areas and aim to shift commuters to public transit (PT) usage \citep{gu2018congestion}. However, such price-based approaches face persistent criticism due to perceived unfairness, as they tend to favor the rich over the poor commuters \citep{arnott1994welfare}; even when compensation is proposed, its adequacy and procedural fairness remain critical concerns \citep{frey1996old}. As an alternative approach that addresses both efficiency and equity, Tradable credit schemes (TCS) are gaining attention for shifting car commuters toward PT \citep{balzer2022modal}, aiming to reduce vehicle traffic volumes and improve system-wide efficiency and equity in multimodal networks.\citep{verhoef1997tradeable,nie2013managing,wu2012design,balzer2022modal}.

TCS encourages a shift from car use to PT by assigning individuals a limited number of credits, making driving more costly than the initial allocation allows, while preserving individual flexibility in route and mode choices through credit trading. Travelers' generalized costs are ultimately balanced through monetary compensation gained by PT users who sell unused credits, and additional monetary costs by drivers who purchase extra credits. This market-based mechanism increases the relative cost of car travel and enhances the attractiveness of PT, thereby promoting a transition toward more sustainable and efficient mobility.

TCS has been widely discussed as a tool for road traffic congestion management \citep{yang2011managing,xiao2013managing,bao2019regulating}. 
Furthermore, a growing literature investigates the impact of TCS in multimodal settings. \citet{wu2012design} utilizes a logit model to capture the travelers' choice options among car, PT, and no-travel, and further develops a user equilibrium model within a Mathematical Program with Equilibrium Constraints (MPEC) framework for system optimal solutions under both congestion pricing and TCS. \citet{nie2013managing} adopted a bottleneck-based morning commute model with two competing options, car with bottleneck versus uncongested transit or carpooling options, highlighting how TCS can shift demand from congested traffic modes. In a similar vein, \citet{tian2013tradable} studied departure time and mode choice equilibria under user heterogeneity and found that implementing TCS led to slight reductions in generalized cost. \citet{chen2023market} investigated TCS market design issues, such as initial credit allocation, expiration rules, transaction fees (fixed vs. proportional), and regulatory interventions, within a two-mode (private car and public transit) system through simulation studies. Later,  \citet{balzer2022modal} introduced a trip-based Macroscopic Fundamental Diagram (MFD) framework \citep{leclercq2017dynamic, mariotte2017macroscopic,lamotte2018morning} that incorporates heterogeneous trip lengths and accumulation-based traffic dynamics. This allowed the model to better capture how peak-hour demand shifts over time in response to TCS policies. Subsequently, \citet{balzer2023dynamic} extended the framework by allowing flexible departure time choices, and adopted the generalized bathtub model \citep{jin2020generalized} to represent network-wide traffic accumulation and the interaction between speeds and congestion under different transportation modes (private cars, public transportation, and carpooling).

Although existing studies above have demonstrated the potential of TCS to foster a mode shift to PT or shared modes, existing models often assume unlimited public transit capacity or fixed occupancy for carpooling, neglect waiting time as a key determinant of mode choice behavior, and treat in-vehicle travel time as a sufficient proxy for the time-based cost. These assumptions overlook the fact that under large-scale modal shifts induced by TCS, public transit services can become saturated, leading to longer waiting times and reduced service quality, which may impact users’ willingness to switch to PT. As a result, this could undermine public acceptance of the TCS and bias the system performance estimation, especially under insufficient bus or shared transport supplies. In response to these challenges, public authority may improve the PT service while implementing TCS. However, achieving an optimal balance often necessitates careful consideration of the trade-offs between traffic performance, cost-effectiveness, and societal impacts.



To address these gaps, we develop a dynamic multimodal network equilibrium model that capture operation constraints from supply side and induced waiting times on demand side under TCS regulation. In contrast to existing studies that assume unlimited capacity and fixed service levels, our model incorporates supply-side dynamics such as vehicle dispatching time, fleet size, and vehicle capacity features, and considers time-varying demand. The proposed framework integrates a trip-based multimodal MFD for traffic congestion dynamics, a point queue model \citep{zhang2013modelling} to characterize station-level waiting time, and a logit-based choice model that links perceived generalized cost to probabilistic mode choices of travelers across available travel alternatives. This integrated structure enables us to analyze the interactions among travelers’ mode choices, traffic dynamics, credit trading behavior, and operational service constraints comprehensively. We formulate the equilibrium condition as a Variational Inequality (VI) problem and discuss its existence and uniqueness accordingly. To solve the model, we propose a gradient projection method with backtracking, leveraging automatic differentiation (using Pytorch) and a principled selection of initial points to improve computational traceability.

This model structure captures the trade-offs among travel time, waiting time, and monetary cost faced by different user groups. A key feature is that the waiting times for bus and DRAS modes are endogenously determined, reflecting realistic service or operation limitations under increased demand, which is particularly important in systems regulated by TCS. However, it assumes a fixed supply of transport resources, and this assumption may not hold in practice, as large-scale demand shifts triggered by TCS may overwhelm existing public transport services and necessitate operational adjustments. To address this supply-side challenge and enhance system performance, we further integrate demand-side regulation through TCS with adaptive supply adjustments. Specifically, we introduce a Demand-Responsive Autonomous Shuttle (DRAS) system as a flexible alternative to bridges the gap between private cars and traditional fixed-schedule public transit. We further propose a bi-level optimization model that jointly optimizes credit allocations and DRAS operations, including fleet sizing and scheduling, offering a coordinated approach to manage modal shifts and improve system efficiency.

To summarize, the main contribution of our paper can be summarized as:
\begin{itemize}
    \item We propose an integrated dynamic equilibrium framework that explicitly captures operational constraints (e.g., the discrete capacity of shared and public transport vehicles) and the induced waiting times faced by travelers. By embedding these mechanisms, the model produces a more realistic representation of user disutility and thereby delivers more credible predictions of multimodal travel behavior under TCS regulation.
    \item We develop a hybrid dynamic framework that combines discrete-time demand decisions with continuous-time service vehicle tracking. While car users are modeled in discrete departure intervals, the operations of buses and DRAS are tracked continuously to capture station-level waiting avoid being overly dependent on the length of the discrete time. 
    \item We adopt a unified speed function across all modes that shares the same road space, rooted in a multimodal MFD framework. This resolves a limitation in existing papers \citep{balzer2022modal,balzer2023dynamic} that each mode has independent speed and congestion dynamics, and enables more realistic feedback between modes competing for the same infrastructure, thereby allowing for the analysis of DRAS operating problems efficiently.
    \item We further integrate the equilibrium loop as a component of the system-level decision process and propose a bi-level optimization model to jointly determine tradable credit allocation and DRAS fleet sizing and dispatching. The proposed design unifies demand regulation and supply adaptation within a single optimization framework, enabling coordinated policy interventions across both sides.   
\end{itemize}

\section{Methodology}

\subsection{Problem Definition}
We consider a multimodal transportation system in a corridor with multiple origin-destination (OD) pairs. Travelers have the option to choose from three travel modes: private cars, fixed-schedule bus, and DRAS. Each mode has unique characteristics in terms of both capacity and flexibility. For instance, private cars generally provide low capacity but high flexibility, DRAS offers medium capacity and moderate flexibility, and conventional buses provide high capacity but operate with relatively fixed routes and schedules. Travelers are assumed to behave rationally, choosing a travel mode, either car, bus, or DRAS, that minimizes their individual generalized cost. These costs include in-vehicle travel time, waiting time, and monetary payments or compensations related to credit trading. These choices made by travelers, in turn, affect traffic congestion levels and waiting times for shared transport modes. We represent this behavioral response through a stochastic user equilibrium framework based on a logit model, where no traveler can reduce their cost by unilaterally switching modes given travelers' departure time, prevailing system conditions, and credit prices.

To mitigate road congestion and promote sustainable travel behavior, we explore management strategies from both the demand side and the supply side. This includes implementing the TCS alongside adjustments in service operations. Under TCS, private vehicle users must purchase extra credit to cover their trips, while travelers who use the bus or DRAS can sell their unused credits. For the service operations, we adjust the fleet size, headway, and departure time to meet the DRAS demand dynamically.  

Overall, this problem involves the behavioral responses of travelers in relation to decisions made by operators and policymakers. This creates a bilevel optimization problem in which the upper level, representing operators or policymakers, optimizes system performance through two intervention levers:

\begin{itemize}
    \item The initial allocation of tradable credits to travelers and the required credits for driving trips.
    \item The sizing and dispatch schedule of the DRAS fleet.
\end{itemize}

These upper-level decisions are formulated in conjunction with the lower-level responses, which encompass travelers' behaviors in relation to their travel decisions and their reactions to the introduced policies and interventions.

\subsection{Modeling Framework}
In this section, we first present the modeling framework for a multimodal transportation system under TCS regulation. Secondly, we introduce the bi-level optimization structure, with the upper level addressing the system operator’s decisions on credit allocation and DRAS operations, and the lower level consisting of the equilibrium model. The equilibrium framework comprises five interconnected components, each capturing a key aspect of system dynamics and user behavior.
\begin{enumerate}
    \item A trip-based multimodal MFD describes the evolution of traffic conditions over the time horizon, capturing congestion effects shared across all modes.

    \item A service vehicle module (i.e. buses and DRAS) characterizes vehicle arrival patterns at stations, subject to capacity and operational constraints, including fleet size, dispatch frequency, and scheduled departure times.

    \item A point queue model estimates passenger waiting times at stations by modeling the interaction between the arrival flows of passengers and service vehicles over time.

    \item A logit-based mode choice model captures travelers’ probabilistic decisions based on total generalized cost, including in-vehicle travel time, waiting time, and monetary costs or compensation from credit trading.

    \item A TCS module controls the allocation and consumption of travel credits, determining how credit trading alters credit price, thereby affecting generalized costs and travel behavior.
\end{enumerate}

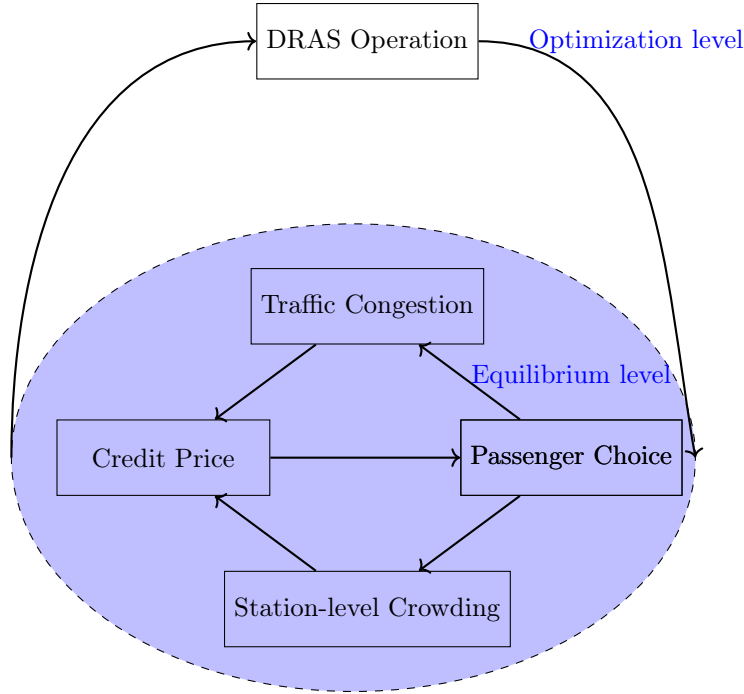
\begin{figure}[!hbtp]
    \centering
\begin{tikzpicture}[
  node distance=1.5cm and 2.5cm,
  every node/.style={font=\small},
  box/.style={draw, minimum width=2.8cm, minimum height=1cm, align=center},
  dashedregion/.style={draw=blue!60, fill=blue!15, dashed, thick, inner sep=0.3cm, rounded corners},
  arrow/.style={->, thick},
]

\coordinate (ellipsecen) at (2.5,0);
\fill[blue!25, draw=black, dashed] (ellipsecen) ellipse (4.5 and 3.1);

\coordinate (ellipseleft) at ($(ellipsecen) + (-4.5, 0)$);
\coordinate (ellipseright) at ($(ellipsecen) + (4.5, 0)$);

\node[box] (credit) at (0,0) {Credit Price};
\node[box, right=of credit] (choice) {Passenger Choice};
\node[box, above=of $(credit)!0.5!(choice)$] (cong) {Traffic Congestion};
\node[box, below=of $(credit)!0.5!(choice)$] (wait) {Station-level Crowding};


\node[box, above=2.5cm of cong, label={[xshift=15pt]right:\textcolor{blue}{Optimization level}}] (dras) {DRAS Operation};

\node[box, right=of credit, label={[yshift=8pt]above:\textcolor{blue}{Equilibrium level}}] (choice) {Passenger Choice};

\draw[arrow] (credit) -- (choice);
\draw[arrow] (choice) -- (cong);
\draw[arrow] (choice) -- (wait);
\draw[arrow] (wait) -- (credit);
\draw[arrow] (cong) -- (credit);

\draw[arrow] (ellipseleft) to[out=90, in=180] (dras.west);
\draw[arrow] (dras.east) to[out=360, in=100] (ellipseright);
\end{tikzpicture}
    \caption{Bi-level framework linking DRAS operation with equilibrium dynamics.}
    \label{fig:overall framework}
\end{figure}

These components interact in a closed feedback loop. Traffic dynamics determine vehicle travel times and influence the scheduling of service vehicles. These operational factors shape users’ experience, particularly through their impact on travel and waiting times. In response, travelers adjust their mode choices and credit trading behavior, which then feed back into mode-specific demand distribution and network conditions, ultimately affecting traffic dynamics again. A user equilibrium is reached when no traveler can improve their outcome by unilaterally changing their travel mode, given the prevailing travel conditions and credit prices. On top of this equilibrium structure, we introduce an outer-layer optimization problem that determines system-level decisions, specifically the sizing of the DRAS fleet. This leads to a bilevel optimization framework, where system-level decisions on service provision are made taking into account the equilibrium responses of users. (see \autoref{fig:overall framework} as a reference)
\

The subsequent subsections present detailed descriptions of the individual modules constituting our framework.

\subsection{Prelimnary settings}
We assume that travel demand is defined in continuous time. However, to balance computational efficiency with modeling resolution, the analysis period is discretized into $M$ intervals, each of length $\Delta t$, with time points $t_0, t_1, t_2, \dots, t_M$, where $t_m = m\Delta t$. We consider $N$ commuter groups indexed by $i$, each with total demand $q_i$, a trip length $l_i$, and a scheduled departure time $d_i$ and $q_i, l_i, d_i \in \mathbb{R}_+$. To simplify the analysis and ensure tractability in an equilibrium setting, travelers are modeled as non-atomic, allowing $q_i(t_m)$ to be treated as a continuous quantity over time. There are three modes available for users to choose from: private car, bus with fixed timing and large capacity, and DRAS with flexible timing and smaller capacity.\\

\begin{itemize}
    \item Private car: Travelers driving private vehicles depart immediately without waiting. Their travel time depends on privailing road congestion. A fixed amount of tradable credits is required per trip, and travelers need to purchase additional credits if their allocation is insufficient.

    \item Bus: This mode represents a traditional, high-capacity public transport system operating on fixed schedules. Travelers choosing this mode may experience waiting at the stations along the path due to fixed dispatches, and in-vehicle travel time is affected by road traffic conditions. Passengers pay a fixed fare and can sell unused travel credits in exchange for compensation.

    \item DRAS: DRAS offers lower-capacity but flexibly dispatched services that adapt to real-time demand within each interval. The dispatch decision is based on accumulated demand at the stations: if a sufficient number of passengers arrive early, the vehicle will depart. As a result, waiting time is endogenously determined by both demand levels and dispatch strategies. Passengers experience both waiting time at stations and in-vehicle travel time affected by road traffic congestion. DRAS charges a fixed fare, and passengers could also sell their allocated credits for monetary compensation.
\end{itemize}

We further assume that every origin or destination has an associated station,  allowing travelers to complete their entire trip using their selected PT modes. Accordingly, we define the decision variables $\mathbf{x}, \mathbf{y}, \mathbf{z} \in [0,1]^M$, where $x_i(t_m), y_i(t_m), z_i(t_m) \in [0,1]$ represent the proportion of travelers departing from origin $i$ in interval $[t_{m-1}, t_m]$ who choose car, bus, and DRAS, respectively. Therefore, we define the decision variables $\mathbf{x}, \mathbf{y}, \mathbf{z} \in [0,1]^M$, where each element $x_i(t_m), y_i(t_m), z_i(t_m) \in [0,1]$ represents the proportion of travelers departing in interval $[t_{m-1}, t_{m}]$ who choose car, bus, and DRAS, respectively. As the credits are traded freely within a regulated market (e.g., credit expiration, time for initial allocation), the decision variable credit price $p \geq 0$ is determined endogenously within the system and is primarily influenced by the initial credit allocation parameter $k$ and the per trip credit requirement $\tau$ for private car use.

Therefore, we have the variable domain for the system is defined as:

\begin{equation}
  K = \left\{ (\mathbf{x}, \mathbf{y}, \mathbf{z}, p) \;\middle|\;
\mathbf{x}, \mathbf{y}, \mathbf{z} \in [0,1]^M,\;
\mathbf{x} + \mathbf{y} + \mathbf{z} \leq \mathbf{1},\;
p \geq 0
\right\}.
\label{early domain}  
\end{equation}

While traveler decisions are modeled in discrete time, we adopt continuous-time tracking for the arrival and departure of buses and DRASs. This prevents the waiting times, which are induced by operational constraints, from being overly dependent on the length of the discrete time interval. For example, if the time step is close to the vehicle headway, small shifts in departure time could lead to large jumps in waiting time due to the discrete time length setting. By using continuous-time tracking for vehicle operations, we avoid this issue and obtain more accurate estimates of passenger waiting times. Beside, this continuous tracking does not significantly increase computational burden due to the relative small amount of service vehicles compared to private vehicles. Details of this hybrid modeling structure are presented in the next section.

\subsection{Trip-based MFD in discrete time horizon}
\label{mfd}
We consider a single-area, trip-based multimodal MFD framework, where each traveler group has a predefined departure time and a fixed travel distance. We adopt a setting in which all travel modes share a single multimodal MFD, reflecting their concurrent use of the same road infrastructure. This unified formulation captures speed dynamics and congestion interactions across modes through a single, consistent speed–flow relationship. As a result, the completion time of a trip is determined by its departure time, the prevailing network mean speed (during circulation in the network), and its travel distance. The prevailing traffic speed varies dynamically with vehicle accumulation \citep{mariotte2017macroscopic, lamotte2018morning}. Therefore, we have the car travel distance $l_i^{\text{car}}$ satisfies:
\begin{equation}
    l_i = \int_{d_i}^{d_i + T_i} V(n(t)) \, dt, \; \text{where, } n(t) := n^{\text{car}}(t) + n^{\text{bus}}(t) + n^{\text{DRAS}}(t)
    \label{continue}
\end{equation}
where $V_{\text{car}}(\cdot)$ denotes the speed as a function of the total accumulation of all vehicle types over time. This formulation links the dynamic network condition to experienced travel time, forming the core of our traffic congestion modeling. 

In a discretized time form, if the travelers' departure $d\in [t_{m-1}, t_{m})$, we assume these travelers enter the network during the same time slot and contribute to the accumulation $n^{car}(t_m)$ of the time slot $t_m$. However, since buses and Demand Responsive Autonomous Shuttles (DRAS) operate with relatively small fleet sizes compared to private vehicles, we assume that they share the same average speed as cars while in operation, but do not contribute to network accumulation. Therefore, the speed at time \(t_m\) is determined solely by the number of private vehicles on the road, denoted as \(n(t_m)\). Consequently, we can approximate the travel distance for cars, \(l^{car}\), as follows:
\begin{equation}
   l_i \approx \sum_{t_m: d_i \in [t_{m-1}, t_{m})}^{t_m + T_i} V(n(t_m)) \cdot \Delta t,\;\text{where, } n(t_m) := n^{\text{car}}(t_m) 
   \label{discretemfd}
\end{equation}
This expression captures the cumulative distance covered by a bunch of vehicles from $t_m$, where its departure time $d_i\in [t_{m-1}, t_{m})$, until the end of its trip duration $T_i$,
based on prevailing traffic speeds. Here, time is divided into discrete intervals $[t_{m-1}, t_m)$, and for each interval, the vehicle is assumed to travel at a constant speed within each discrete intervals $[t_{m-1}, t_m)$, given by the speed function $V(n(t_m))$, which depends on the number of cars $n^{\text{car}}(t_m)$ present in the system during that time slot. The total distance is then obtained by summing speed times time step $\Delta t$ over all relevant intervals. This discrete-time trip-based multimodal MFD aligns with the aggregate nature of MFD models, but focuses on cumulative flows rather than individual trajectories.


As vehicles' accumulation $n^{\text{car}}(t_{m})$ evolves over discrete time, it follows the dynamic update equation:
\begin{equation}
 n^{\text{car}}(t_{m}) = n^{\text{car}}(t_{m-1}) + q_{\text{in}}(t_m) - q_{\text{out}}(t_m),   
\end{equation}
where $q_{\text{in}}(t_m)$ and $q_{\text{out}}(t_m)$ denote the number of car users entering and exiting the network during time slot $t_m$, respectively.

Given a set of traveler groups $i$, each defined by a departure time $d_i$ and group size $q_i(t)$, the number of car users entering the network at time $t_m$ is:
\begin{equation}
    q_{\text{in}}(t_m) = \sum_{i: d_i \in [t_m, t_{m+1})} x_i(t_m) \cdot q_i(t_m)
\end{equation}
where $q_i(t_m)$ is the total number of demand within group $i$ departing at time $t_m$.

For the exit flow $q_{out}$, we adopt virtual travelers, by referring to \citep{lamotte2018morning,balzer2022modal,balzer2023dynamic}, for each group $t \rightarrow z(t)$ to keep track of cumulative traveled distance, and also define $t \rightarrow n(t)$ to track the accumulation as a function of the traveled distance, where track the number of active trips with reaming distance higher than zero at t. Therefore the travel time $T_i$ of group $i$ can be:
\begin{equation}
     T_i = z^{-1}(z(d_i) + l_i)
    \label{finishtime}
\end{equation}

Although the model operates over a discretized time horizon, we adopt a continuous-time format for the calculation of arrival and departure times of buses and DRAS. Therefore, the in-vehicle travel time for a service vehicle following the continuous time relationship \autoref{continue}. We adopt a piecewise-constant approximation of speed over each discrete interval, and align departure times within their containing intervals. For example, a DRAS departing at $t = 2$ is assumed to experience the average speed over the entire $[0,5)$ interval. Suppose this traveler travels for $20$ minutes (until $t = 22$),  the travel distance and travel time would be, 
\begin{align}
    l^{\text{DRAS}} = \int_{2}^{22} V(n^{\text{car}}(t)) \, dt  &=  \int_{2}^{5} V(n^{\text{car}}(t)) \, dt + \int_{5}^{20} V(n^{\text{car}}(t)) \, dt + \int_{20}^{22} V(n^{\text{car}}(t)) \, dt\\
    & \approx (5 - 2) \cdot V(n^{\text{car}}(5)) \\&+ \sum_{t_m \in \{[5,10),[10,15),[15,20)\}}  \Delta t \cdot V(n^{\text{car}}(t_m)) + (22 - 20) \cdot V(n^{\text{car}}(20))
\end{align}

\begin{figure}[h]
    \centering
    \includegraphics[width=1\linewidth]{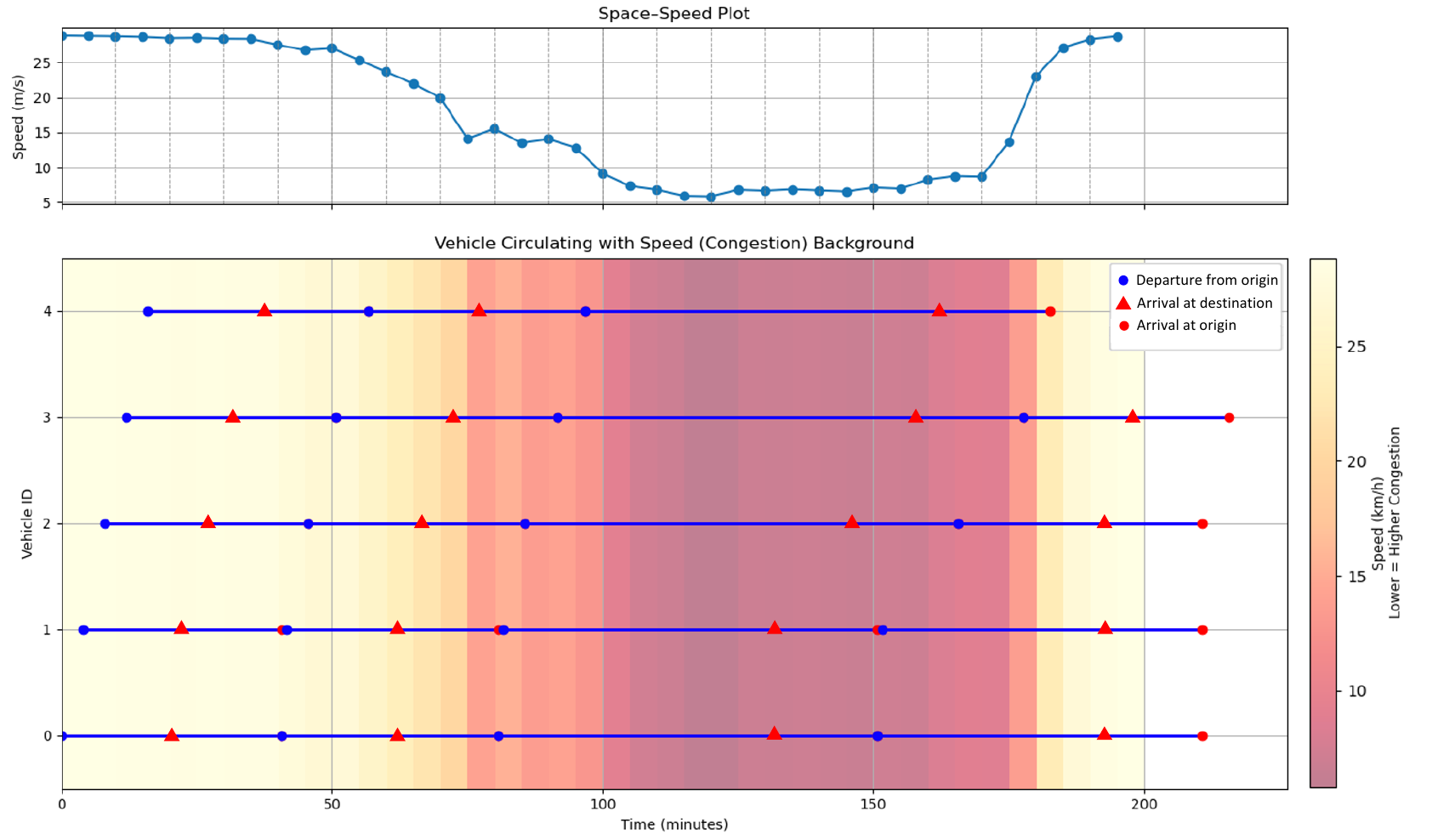}
    \caption{An example of five DRAS schedules in a single OD scenario}
    \label{fig:loop}
\end{figure}

See \autoref{fig:loop} for an example. In this case, five DRAS vehicles are scheduled to operate on a single OD pair over a 200-minute horizon, with staggered departures starting at 4-minute intervals and a minimum headway of 1 minute. The top plot shows the evolution of mean traffic speed over time, while the bottom plot illustrates the dispatching patterns of the five DRAS vehicles as they travel from their origins to the destination and return to their origins. 

We assume that each unit of travel demand corresponds to a single individual. Under this assumption, the passenger flow is equal to the vehicular flow for private car users. However, for buses and DRAS, the relationship between passenger flow and vehicle flow depends on vehicle capacity, as these vehicles can serve multiple travelers. Because we model demand as a continuous variable, we do not track individual loading or vehicle-level matching. Instead, we capture the interaction between vehicle supply and passenger demand at an aggregate level, based on available seat capacity and scheduled or responsive departures. The next section will elaborate on this.

\subsection{Point queue model for station-based waiting}
\label{pointqueuesection}
To capture pre-boarding waiting times, we adopt a point queue model at each station that tracks the mean waiting time over discrete time intervals. The waiting time at each station is jointly determined by the arrival patterns of passengers, as well as the arrival and departure patterns of buses and DRASs, along with their available capacities. In cases where passenger overflow occurs (i.e., when not all waiting passengers can be served by the arriving vehicle), those unserved passengers remain in the queue and are carried over to the next available service. This mechanism allows us to simulate both in-vehicle travel time and station-level waiting time, thereby enabling an evaluation of TCS effects not only on travel time but also on the temporal accessibility of public or shared transportation under TCS-induced mode shifts. More importantly, we explicitly account for how demand fluctuations impact the availability of alternative modes, particularly through public transport waiting times, when solving for the user equilibrium.

\subsubsection*{DRAS or Bus demand}
In our setup, we denote $s$ as the index for stations in the network.  A single station $s$ may serve multiple groups $i$, for example, groups departing from the same origin at different times, traveling to different destinations, so having different trip lengths. Travelers choose their mode, car, bus, or DRAS, instantly when they enter the network, and keep using the same mode until their destination.

\begin{figure}[h]
\centering
\includegraphics[width=1\linewidth]{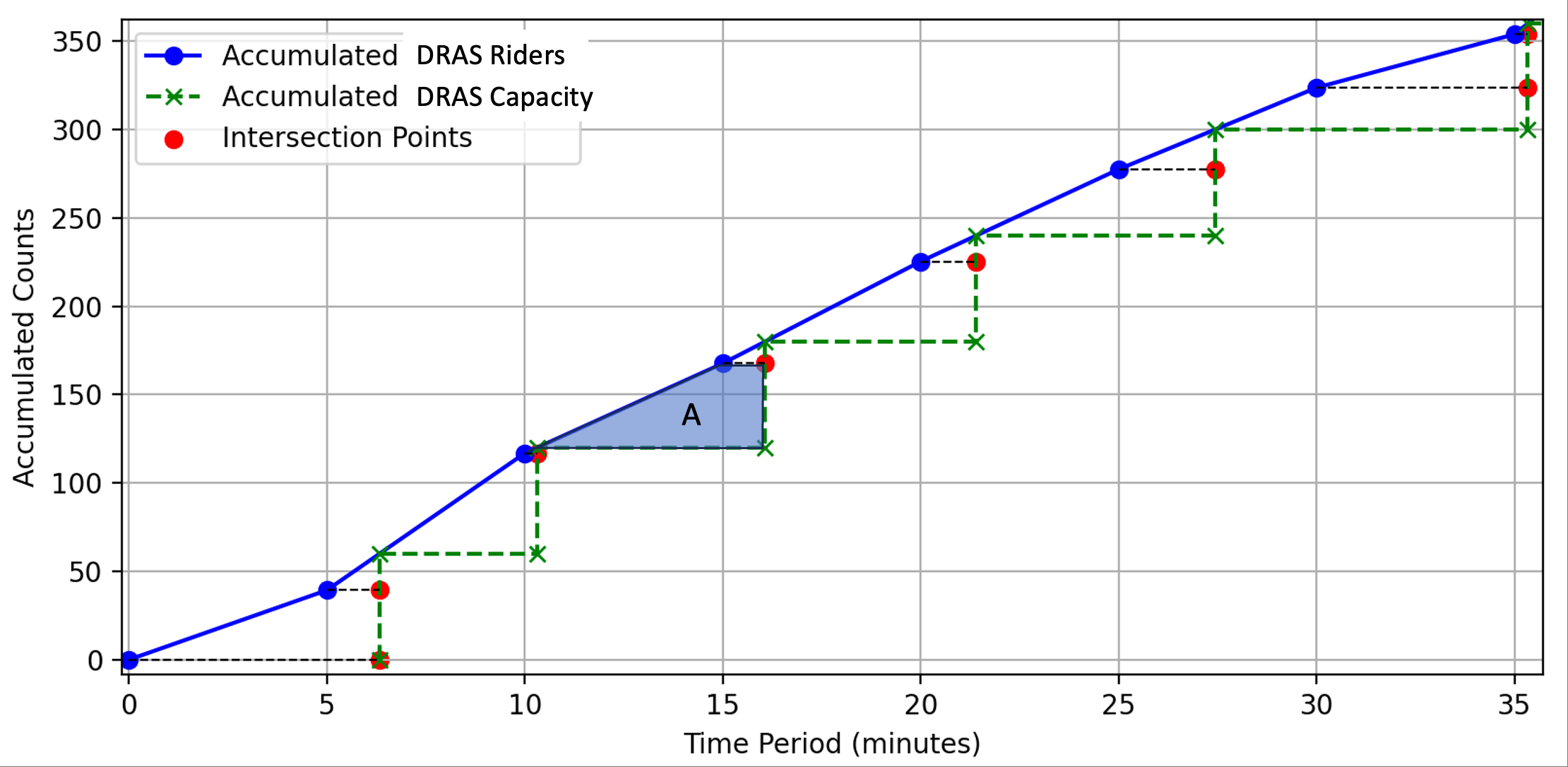}
\caption{Cumulative DRAS arrival and service curves illustrating passenger waiting times.}
\label{example}
\end{figure}

We further assume that DRAS or bus passengers arrive uniformly within each time interval $[t_{m-1}, t_{m})$ at a station $s$, which is depending on the total demand $q_s(t_m)$ and ratio of choosing DRAS $z_s(t_m)$ or buses $y_s(t_m)$ for travel within time $t_m$. Taking DRAS as an example, the cumulative passenger arrival curve at station $s$ is defined as:
\begin{equation}
   A^{\text{DRAS}}_s(t) = A^{\text{DRAS}}_s(t_m) + q_s(t_m)\cdot z_s(t_m) \cdot (t - t_m), \quad \text{for } t \in [t_{m-1}, t_{m}], \quad m \in \mathbb{Z}^+, \ m \ge 1
   \label{demand_cruve}
\end{equation}
with $A^{\text{DRAS}}_s(t_0) = 0$ as the initial condition. 
Here, $A^{\text{DRAS}}_s(t)$ denotes the cumulative number of DRAS passengers arriving at station $s$ by time $t$, where $t \in [t_{m-1}, t_m]$ and $m \in \mathbb{Z}_{\geq 1}$. 
We let $\mathcal{I}(s)$ be the set of traveler groups departing from origin station $s$, and define
\[
q_s(t_m) = \sum_{i \in \mathcal{I}(s)} q_i(t_m),
\]
where $q_i(t_m)$ is the demand of group $i$ in interval $[t_{m-1}, t_m]$.  
Thus, $q_s(t_m)$ represents the total demand at station $s$ during this interval, 
and $z_s(t_m) \in [0,1]$ is the decision variable denoting the proportion of these travelers choosing DRAS.  
This arrival process is illustrated in \autoref{example}, which shows the cumulative passenger arrival curves and the corresponding DRAS service vehicle curves.

\subsubsection*{DRAS operation and supply curve}
To model DRAS operations, we define three key parameters: the total fleet size $N_{\text{DRAS}}$, the $n$th departure time $t_{n}^{\text{DRAS},j}$ of the shuttle $j$, and the vehicle capacity $C_{\text{DRAS}}$ (assume to be universal for all operating shuttles). These parameters jointly determine the operations of a fleet of shuttles circulating along a fixed route. As each vehicle repeatedly visits stations $s$ over time, its successive departures collectively generate the cumulative service vehicle capacity at that station. The resulting service profile is thus represented as a cumulative capacity curve reflecting the supply of DRAS seats available to passengers over time (see the dashed green lines in \autoref{example} for example).

We assume that buses and DRAS vehicles share the same set of visited locations (not necessarily the same stations) for passenger pickup and drop-off. At each station $s$, a given shuttle $j$ may visit multiple times during the modeling horizon. Therefore, we denote its $n$th arrival and departure times at station $s$ as $t_{s,n}^{\text{arr},j}$ and $t_{s,n}^{\text{dep},j}$, respectively. As a DRAS, the shuttle departs when a minimum occupancy threshold is reached. Let $\omega \in (0,1]$ be the minimum sufficient occupancy threshold ratio. Then the departure time is given by:
\begin{equation}
    t_{s,n}^{\text{dep},j} = 
\inf\left\{ t \ge t_{s,n}^{\text{arr},j} \;\middle|\; A_s^{\text{DRAS}}(t) - A_s^{\text{DRAS}}(t_{s,n}^{\text{arr},j}) \ge \omega C_{\text{DRAS}} \right\}\;
\end{equation}
where $A_s^{\text{DRAS}}(t)$ denotes the cumulative number of DRAS boarding requests at station $s$ by time $t$. The above equation captures earliest time when the occupancy threshold is first met. $A_s^{\text{DRAS}}(t) - A_s^{\text{DRAS}}(t_{s,n}^{\text{arr},j})$ is the number of passengers that have been waiting since the vehicle arrived. In practice, a shuttle may find that sufficient demand has already been accumulated when it arrives at the station, particularly during high-demand periods, and no passengers get off. In such cases, the departure condition may be satisfied immediately, and the vehicle departs without any additional waiting. This case is implicitly captured by the formulation above, where the threshold condition is evaluated starting from the arrival time.

Overall, the corresponding cumulative capacity curve of shuttle $j$ at station $s$, during its $n$th visit, is defined as:
\begin{equation}
   S_{s,n}^{\text{DRAS},j}(t) =
\begin{cases}
S_{s,n}^{\text{DRAS},j}(t_{s,n}^{\text{arr},j}), & \text{for } t \in [t_{s,n}^{\text{arr},j},\; t_{s,n}^{\text{dep},j}) \\
S_{s,n}^{\text{DRAS},j}(t_{s,n}^{\text{arr},j}) + \omega C_{\text{DRAS}}, & \text{for } t = t_{s,n}^{\text{dep},j}
\end{cases} 
\end{equation}
where, $S_{s,n}^{\text{DRAS},j}(t)$ is the cumulative number of passengers served by vehicle $j$ at station $s$ during its $n$th visit, up to time $t$. $t_{s,n}^{\text{arr},j}$, $t_{s,n}^{\text{dep},j}$ are the arrival and departure times of the vehicle at the station, $A_s^{\text{DRAS}}(t_{s,n}^{\text{dep},j}) - A_s^{\text{DRAS}}(t_{s,n}^{\text{arr},j})$ is the number of passengers that arrived during the dwell time. This formulation illustrates that the cumulative service curve remains flat while the vehicle is waiting at the station, and jumps at the departure time to reflect the number of passengers served, which is bounded by the occupancy threshold.

Let $S_{s}^{\text{DRAS},j}(t)$ denote the cumulative service curve of shuttle $j$ at station $s$, which records the total number of passengers that vehicle $j$ has served up to time $t$. Since all shuttles operate along the fixed routes, their contributions to station-level capacity are aligned along the timeline. The total cumulative service curve at station $s$, denoted $S_{s}^{\text{DRAS}}(t)$, is obtained by summing the vehicle-level service curves pointwise in time:
\begin{equation}
 S_{s}^{\text{DRAS}}(t) = \sum_{j=1}^{N_{\text{DRAS}}} S_{s}^{\text{DRAS},j}(t)   
\end{equation}
Each curve $S_{s}^{\text{DRAS},j}(t)$ reflects the actual service provided by vehicle $j$ at each point in time, and the sum gives the total number of passengers that have been served at station $s$ by all DRAS vehicles up to time $t$.

To ensure the feasibility of the schedule and avoid unrealistically tight departures, we impose a minimum headway constraint between consecutive DRAS departures at each station. Therefore, for any two consecutive departures \( t_{s,{n-1}}^{\text{dep}} \) and \( t_{s,n}^{\text{dep}} \), the following condition must hold:
\begin{equation}
t_{s,n}^{\text{dep}} - t_{s,{n-1}}^{\text{dep}} \geq H_{\min}
\end{equation}
where \( H_{\min} \) denotes the minimum allowable time interval between two successive departures from the same station. The departure times $t_{s,{n-1}}^{\text{dep}}$ and $t_{s,n}^{\text{dep}}$ correspond to the jump points on the cumulative service curve $S_s^{\text{DRAS}}(t)$, where the curve increases due to an DRAS departure.

\subsubsection*{Bus operations}
Bus operations differ from DRAS in that buses follow a fixed departure schedule. Their departure times do not change based on real-time passenger demand at stations. Although the arrival curve for bus passengers is constructed in the same way (following \autoref{demand_cruve}), which is based on the share of demand choosing the bus for travel, denoted $y_s(t_m)$, the actual departure process is determined by a predefined timetable, and fixed dwell times at each stop.

Let $T_{\text{dep}}^{\text{bus}} = \{t_1^{\text{bus}}, t_2^{\text{bus}}, \dots, t_N^{\text{bus}}\}$ be the set of scheduled bus departure times at a given station $s$. At each time $t_n^{\text{bus}}$, the bus departs with all passengers currently waiting, up to the vehicle’s available capacity $C_{s,\text{bus}}$ at station $s$. Therefore, the number of passengers boarding at the prevailing time is the smaller of the available demand $y(t_n^{\text{bus}})$ or the bus available capacity. The cumulative bus service curve, denoted $ S^{\text{bus}}_s(t)$, represents the total number of passengers who have departed by bus up to time $t$ at station $s$, and is defined as:
\begin{equation}
  S^{\text{bus}}_s(t) = 
\begin{cases} 
S^{\text{bus}}_s(t_{n-1}^{\text{bus}}), & \text{if } t \in [t_{n-1}^{\text{bus}},\; t_n^{\text{bus}}), \\
S^{\text{bus}}_s(t_{n-1}^{\text{bus}}) + \min\left\{y_s(t_n^{\text{bus}}),\; C_{s,\text{bus}} \right\}, & \text{if } t = t_n^{\text{bus}},
\end{cases}  
\end{equation}
where, $S^{\text{bus}}_s(t)$ is the cumulative number of passengers who have departed by bus up to time $t$. $t_n^{\text{bus}}$ is the $n$th scheduled departure time. $t_{n-1}^{\text{bus}}$ is the previous scheduled departure time. $y_s(t_n^{\text{bus}})$ is the decision variable. $C_{s,\text{bus}}$ is the available capacity at station $s$. The term $\min\left\{y_s(t_n^{\text{bus}}),\; C_{s,\text{bus}} \right\}$ ensures that the number of passengers boarding does not exceed either the waiting demand or the bus capacity. This setup reflects the fixed-schedule nature of bus service: passengers only depart at scheduled times, and the number who can board depends on how many riders are waiting and how many seats are available. On the other hand, their in-vehicle travel time varies based upon the traffic dynamics following equation \autoref{finishtime}.

\subsubsection*{Passenger waiting time}
Passenger waiting time is defined as the temporal gap between the passenger's arrival and the bus or DRAS service (i.e., departure) curves at a given station (see highlighted area A in figure 2 as an example). Similar curve-based methods have been studied intensively in morning commute problems \citep{vickrey1969congestion,daganzo2001simple}, where the same logic was applied to estimate delay in travel times rather than station waiting times. Let $A_s^\mu(t)$ and $S_s^\mu(t)$ denote the cumulative passenger arrival and departure curves for mode $\mu \in \{\text{bus}, \text{DRAS}\}$ at station $s$. For the cumulative total waiting time over a time interval  $[t_{m-1}, t_{m}]$, we have the area between the arrival and departure curves corresponds to the total waiting time.

We define the inverse functions of the cumulative curves as follows:
\begin{itemize}
    \item $t_{\text{arr},s}^\mu(\Delta)$: the time at which the $\Delta$th passenger arrives at station $s$ for mode $\mu$,
    \item $t_{\text{dep},s}^\mu(\Delta)$: the time at which the $\Delta$th passenger departs via mode $\mu$ from station $s$.
\end{itemize}
Then, the total waiting time of passengers arriving during interval $[t_{m-1}, t_{m}]$ (indicated by area A in \autoref{example}) is given by:
\begin{equation}
   W_s^\mu(t_m) = \int_{A_s^\mu(t_{m-1})}^{A_s^\mu( t_{m})} \left( t_{\text{dep},s}^\mu(\Delta) - t_{\text{arr},s}^\mu(\Delta) \right) d\Delta 
\end{equation}


Assuming arrivals are uniformly distributed over the interval, the average waiting time for mode $\mu$ at station $s$ in interval $[t_{m-1}, t_{m}]$ is:
\begin{equation}
 AW_s^\mu(t_m) = \frac{W_s^\mu(t_m)}{A_s^\mu( t_{m}) - A_s^\mu( t_{m-1})}   
\end{equation}
where the denominator is the number of passengers who arrived during the interval.

\subsection*{Perceptive average waiting bias}
When modeling the average waiting time $AW_s^\mu(t)$, the standard approach assumes equal weight for all passengers. However, this approach introduces bias, as longer delays, which typically carry higher psychological or economic burdens, are not adequately captured, leading to a systematic underestimation of perceived disutility \citep{fan2016waiting,ji2019waiting}, even in cases where basic amenities (e.g., timetable board and chairs) are provided.

To address this, we introduce a time-dependent dynamic weighting mechanism to the average waiting time. The idea is to weigh waiting time more heavily when passengers wait longer, particularly in high-demand periods when operational constraints cause waiting spillover effects. We adjust the perceived waiting time $PAW_s^\mu(t)$ based upon the number of users since by assuming the uniform distribution of passengers arrival, the more passengers meaning the longer waiting time. Therefore, we have 
\begin{align}
      PAW_s^\mu(t_m) &= AW_s^\mu(t_m)\cdot \eta\big(A_s^\mu(t_m) - A_s^\mu(t_{m-1})\big)\\
      &=\frac{W_s^\mu(t_m)}{A_s^\mu(t_m) - A_s^\mu(t_{m-1})} \cdot \eta\big(A_s^\mu(t_m) - A_s^\mu(t_{m-1})\big) = \eta W_s^\mu(t_m), 
      \label{paw}
\end{align}
where, $\eta >0$ is the factor of weight, $W_s^\mu(t_m)$ is the standard (unweighted) waiting time. This adjustment ensures that long delays experienced are appropriately emphasized. For example, if only one user arrives in the interval, the weight defaults to 1, and the adjusted waiting time equals the actual individual waiting time. As the number of users increases, the weighting function amplifies the aggregate delay to reflect the broader impact on user experience.


\subsection*{Integration of waiting time and travel time for DRAS}
To account for the feedback loop between service vehicle waiting times and their future arrival times, we adopt an event-based sequential update scheme for computing shuttle arrival and departure times across stations.

The procedure begins by computing each shuttle’s initial arrival times using free-flow travel speeds. Then, we iteratively update the departure times by simulating vehicle service at each station in chronological order. For each arrival event, we calculate the dwell time based on cumulative passenger arrivals and minimum headway rules, and use this to update the vehicle’s subsequent travel schedule.

Each vehicle's timeline is propagated forward sequentially, accounting for updated departure times at each stop. If the computed arrival time for the next stop exceeds the simulation horizon when no vehicles circulating in the network, the process terminates. This ensures consistency between vehicle arrival curves, station dwell times, and resulting travel durations.

\subsection{TCS setting}
As previously mentioned, we consider a static setting for the TCS. Each commuter is allocated an initial credit amout of $k \in \mathbb{N}$ credits at the beginning of the study period. The study period can be a day or a specific peak-hour window (e.g., morning or afternoon). Travelers who choose to drive a private car must pay $\tau \in \mathbb{N}$ credits per car trip. In contrast, those who choose public or shared transit modes are not charged any credits and may sell their unused credits in exchange for monetary compensation.

By setting the required credits for driving $\tau$ higher than the initial allocation $k$, i.e., $\tau \geq k$, drivers are incentivized to purchase additional credits, which allows the regulator to directly control the proportion of car users through the ratio $k/\tau$. Credits are only valid within the study period, ensuring that commuters are motivated to either use or trade them. The unit credit price $p$ is determined endogenously by supply and demand and reflects the marginal willingness to pay for the right to drive.

\subsubsection*{Market clearing condition}
\label{TCS}
A crucial feature of the credit trading market is the market-clearing condition (MCC) \citep{nagurney2000marketable,yang2011managing,balzer2022modal}, which demonstrates the condition that all credits are used by the end of the period. This guarantees that no more credit than the total in the system is consumed when the price is positive, which is a critical outcome to the functioning of the system. If demand for driving exceeds supply, the credit price increases; if supply exceeds demand, prices fall, potentially to zero.

As $q_i$ denote the total demand from commuter group $i$, the total number of credits allocated at the beginning of the period is $\sum_{i=1}^{N} q_i k$, while the total credits consumed by drivers is $\sum_{i=1}^{N} q_i(t)x_i(t)\tau$, where $x_i(t)$ is the share of drivers at time $t$. The MCC can be expressed as the following complementarity relationship:

$$
\begin{cases}
\sum_{i=1}^{N} q_i k = \sum_{t_1}^{t_M} \sum_{i=1}^{N} q_i(t) x_i(t) \tau \quad \Rightarrow \quad p > 0, \\
\sum_{i=1}^{N} q_i k > \sum_{t_1}^{t_M} \sum_{i=1}^{N} q_i(t) x_i(t) \tau \quad \Rightarrow \quad p = 0.
\end{cases}
$$
$$
\begin{cases}
 q k = \sum_{t_1}^{t_M}  q(t) x(t) \tau \quad \Rightarrow \quad p > 0, \\
 q k > \sum_{t_1}^{t_M}  q(t) x(t) \tau \quad \Rightarrow \quad p = 0.
\end{cases}
$$
This condition ensures that when credit demand matches supply, the market clears with a positive price $p$. 

\subsection{Utility function and Logit-Based Mode Choice}

In our framework, the generalized disutility perceived by travelers is modeled as a combination of time-based and monetary components. The time-based component includes in-vehicle travel time, which reflects network congestion dynamics (see Section~\ref{mfd}), and waiting time at stations, which captures operation mechanisms and associated queuing effects. (see Section~\ref{pointqueuesection}). On the other hand, the monetary component incorporates direct travel fares as well as payments or revenues from credit trading under the TCS. 

Let $C_i^\mu(t_m)$ denote the deterministic generalized cost perceived by a traveler in group $i$ when choosing mode $\mu \in \{\text{Car}, \text{Bus}, \text{DRAS}\}$ at discrete time $t_m$. The costs for each mode are given by:
\begin{equation}
\begin{aligned}
    C_i^{\text{Car}}(t_m) &= \alpha T_i^{\text{Car}}(t_m) + (\tau - k)p + \delta_{\text{Car}}, \\
    C_i^{\text{Bus}}(t_m) &= \alpha T_i^{\text{Bus}}(t_m) + \alpha' \sum_{s\in S_{od}} \text{PAW}_s^{\text{Bus}}(t_m) - kp + \delta_{\text{Bus}}, \\
    C_i^{\text{DRAS}}(t_m) &= \alpha T_i^{\text{DRAS}}(t_m) + \alpha' \sum_{s\in S_{od}} \text{PAW}_s^{\text{DRAS}}(t_m) - kp + \delta_{\text{DRAS}}.
\end{aligned}
\end{equation}
Here, $T_i^\mu(t_m)$ is the in-vehicle travel time, $\text{PAW}_s^\mu(t_m)$ is the perceived average waiting-time disutility at station $s$, and $p$ is the equilibrium credit price under the TCS, as introduced in Sections~\ref{mfd}, \ref{pointqueuesection}, and \ref{TCS}, respectively. $\alpha$ and $\alpha'$ are parameters that capture the sensitivities to in-vehicle time and waiting time; $\delta_\mu$ is a mode-specific constant reflecting fixed preferences or penalties such as parking, bus fare, or DRAS fare; $\tau$ and $k$ are the per-trip credit requirement and initial credit allocation.

We then define the random utility for traveler group $i$ and mode $\mu$ as:
\begin{equation}
    U_i^\mu(t_m) = C_i^\mu(t_m) + \varepsilon_i^\mu(t_m),
\end{equation}
where $C_i^\mu(t_m)$ is the deterministic component described above, and $\varepsilon_i^\mu(t_m)$ denotes unobserved factors.

To model this mode choice behavior, we adopt a multinomial logit (MNL) model. Assuming $\varepsilon_i^\mu(t_m)$ are i.i.d. Gumbel-distributed, the probability that a traveler in group $i$ selects mode $\mu$ at time $t_m$ is:
\begin{equation}
    P_i^\mu(t_m) = \frac{\exp(-\theta C_i^\mu(t_m))}{\sum_{\mu' \in \{\text{Car}, \text{Bus}, \text{DRAS}\}} \exp(-\theta C_i^{\mu'}(t_m))},
\end{equation}
where $\theta$ is the scale parameter representing sensitivity to cost differences.

We use the logit model because, in our context, cars, buses, and DRAS operate on the same road infrastructure, so in-vehicle travel times are largely similar across modes and well understood by commuters from their day-to-day experience; hence, they contribute little to the unobserved utility. The dominant source of uncertainty and heterogeneity is the waiting time, which arises from fundamentally different operating schemes. Buses operate on fixed schedules with specific capacities, while DRAS employs a more flexible, demand-responsive approach. These distinct operating processes generate mode-specific waiting-time errors that are approximately independent, which limits the risk of material IIA violations. For analytical tractability and for integration with our dynamic equilibrium and bi-level optimization, we therefore use the standard MNL formulation for mode choice behavior within our dynamic equilibrium framework.

\subsection{Equilibrium Modeling}
\label{equlibrium}
\begin{figure}[h]
    \centering
    \begin{tikzpicture}[node distance=2.5cm and 3cm]
    \node[box] (credit) {Credit\\Price};
    \node[box, right=2cm of credit] (choice) {Commuter Choices};
    \node[box, right=of choice] (queue) {Waiting time\\Point Queue Model};
    \node[box, below=of choice] (mfd) {MFD for Travel Time Estimation
};
    \node[box, below=of queue] (operation) {DRAS \& Bus Operation};
    
    \draw[arrow] (credit) -- (choice) node[near end, above] {\( p \)};
    \draw[arrow] (choice) -- (queue) node[near end, above] {\( \textbf{y}, \textbf{z} \)};
    \draw[arrow] (queue) -- (choice) node[near end, below] {\( PAW_s^\mu(t) \)};
    \draw[arrow] (mfd) -- (operation) node[near end, above] {\( T^{\text{bus}},T^{\text{DRAS}} \)};
    \draw[arrow] (mfd) -- (choice) node[near end, left] {\( T^{\text{car}} \)};
    \draw[arrow] (choice) -- (mfd) node[near end, right] {\( \textbf{x} \)};
    \draw[arrow] (choice) -- (credit) node[near end, below] {\( \textbf{x} \)};
    \draw[arrow] (queue) -- (operation) node[near end, right] {\(  A^{\text{DRAS}}_s(t) \)};
    \draw[arrow] (operation) -- (queue) node[near end, left] {\( S_{s}^{\text{DRAS}}(t),S^{\text{bus}}_s(t) \)};
    \end{tikzpicture}
    \caption{ Interactions of major model components of TCS.}
    \label{fig:structure}
\end{figure}
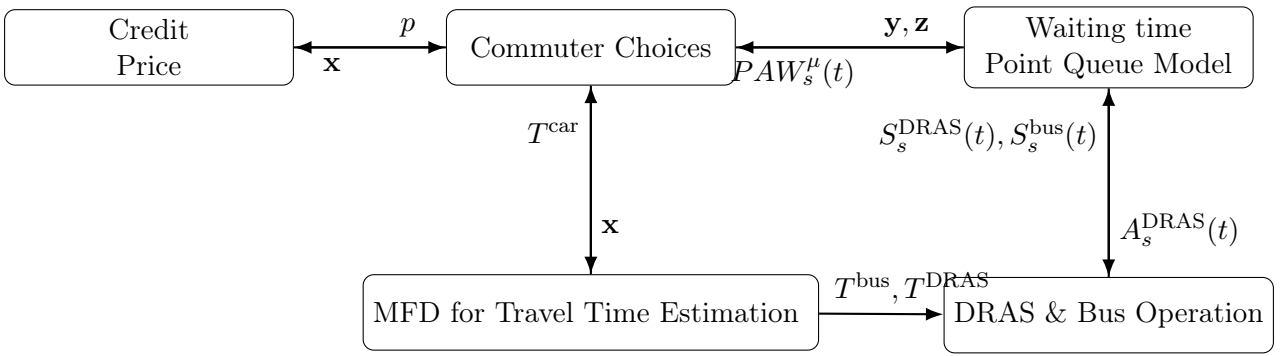

To better illustrate the interactions among the different modules, we present an \autoref{fig:structure} providing a graphical depiction of the connections among traffic dynamics, station-based waiting, service vehicle operation, commuters' choices, and TCS. 

Note that variables on arrows denote inputs to the module at the arrow’s tip.

We now formalize the equilibrium conditions that link these interdependent components. 
For notational simplicity, we omit the group index $i$ and present the model at the aggregate level, 
where all variables represent time-dependent mode shares over the total demand. 
The key decision variables are:

\begin{itemize}
  \item $\mathbf{x} = (x(t_1), \dots, x(t_M))^\top \in [0,1]^M$: share of travelers choosing bar in each time period.
  \item $\mathbf{y} = (y(t_1), \dots, y(t_M))^\top \in [0,1]^M$: share of travelers choosing bus in each time period.
  \item $\mathbf{z} = (z(t_1), \dots, z(t_M))^\top \in [0,1]^M$: share of travelers choosing DRAS in each time period.
  \item $p \in \mathbb{R}_+$: equilibrium credit price in the TCS.
\end{itemize}

The deterministic generalized cost vectors for each mode are:
\[
\mathbf{C}_{\text{Car}} = \big(C_{\text{Car}}(t_1), \dots, C_{\text{Car}}(t_M)\big)^\top,\]
\[
\mathbf{C}_{\text{Bus}} = \big(C_{\text{Bus}}(t_1), \dots, C_{\text{Bus}}(t_M)\big)^\top, 
\]
\[
\mathbf{C}_{\text{DRAS}} = \big(C_{\text{DRAS}}(t_1), \dots, C_{\text{DRAS}}(t_M)\big)^\top.  
\]

Under the multinomial logit specification, the mode choice probability vector at time $t_m$ is:
\[
\mathbf{P}(t_m) =
\begin{pmatrix}
P_{\text{Car}}(t_m) \\
P_{\text{Bus}}(t_m) \\
P_{\text{DRAS}}(t_m)
\end{pmatrix}
=
\frac{\big(\exp(-\theta\, C_{\text{Car}}(t_m)),\; \exp(-\theta\, C_{\text{Bus}}(t_m)),\; \exp(-\theta\, C_{\text{DRAS}}(t_m))\big)^\top}
{\sum_{\mu \in \{\text{Car}, \text{Bus}, \text{DRAS}\}} \exp(-\theta\, C_\mu(t_m))},
\]
where $\theta > 0$ is the scale parameter.

Vectorizing over the full horizon, we have:
\[
\mathbf{P} = \Phi\big( \mathbf{C}(\mathbf{x}, \mathbf{y}, \mathbf{z}, p) \big),
\]
where $\Phi(\cdot)$ applies the logit mapping independently to each $t_m$.

Recall the proposed system operates as a closed feedback loop: mode shares $(\mathbf{x}, \mathbf{y}, \mathbf{z})$ determine traffic conditions and service operations (e.g., DRAS and bus frequencies and departure times), which in turn affect in-vehicle travel times and station-level waiting times. Combined with the credit price $p$, these time-based factors yield the generalized costs $\mathbf{C}$; the generalized costs then update mode shares via $\Phi(\cdot)$, completing the loop.

A user equilibrium is reached when no traveler can reduce their expected disutility by unilaterally switching modes. In this framework, this corresponds to a fixed point where the realized decision vector matches the choice probabilities implied by the generalized costs:
\begin{equation}
 (\mathbf{x}, \mathbf{y}, \mathbf{z}, p) = \Phi\big(\mathbf{C}(\mathbf{x}, \mathbf{y}, \mathbf{z}, p)\big).
\end{equation}

We formulate this fixed-point problem as a Variational Inequality (VI).  
Let $\Phi^\mu(t)$ denote the logit-predicted choice probability for mode $\mu \in \{\text{Car}, \text{Bus}, \text{DRAS}\}$ at time $t$.  
The mapping $F$ is defined as the difference between the realized mode shares and these predicted probabilities:

\begin{equation}
    F_t =
\begin{pmatrix}
x(t) - \Phi^{\text{Car}}(t) \\
y(t) - \Phi^{\text{Bus}}(t) \\
z(t) - \Phi^{\text{DRAS}}(t)
\end{pmatrix},
\quad t = 1, \dots, M,
\end{equation}
where each $\Phi^\mu(t)$ is evaluated at 
$\big(C_{\text{Car}}(t),\, C_{\text{Bus}}(t),\, C_{\text{DRAS}}(t)\big)$.

For the TCS, the MCC requires the total allocated credits equal the total credits consumed by car users, given a positive credit price $p$:
\begin{equation}
 F_p = k - \tau \sum_{t=1}^M \Phi^{\text{Car}}(t).
\end{equation}

Collecting all components, the full mapping $F(\mathbf{x}, \mathbf{y}, \mathbf{z}, p)$ and $(\mathbf{x}, \mathbf{y}, \mathbf{z}, p)$ are:
\[
F(\mathbf{x}, \mathbf{y}, \mathbf{z}, p) =
\begin{pmatrix}
x(t_1) - \Phi^{\text{Car}}(t_1) \\
\vdots \\
x(t_M) - \Phi^{\text{Car}}(t_M) \\
y(t_1) - \Phi^{\text{Bus}}(t_1) \\
\vdots \\
y(t_M) - \Phi^{\text{Bus}}(t_M) \\
z(t_1) - \Phi^{\text{DRAS}}(t_1) \\
\vdots \\
z(t_M) - \Phi^{\text{DRAS}}(t_M) \\
k - \tau \sum_{t=1}^M \Phi^{\text{Car}}(t)
\end{pmatrix},
\quad
(\mathbf{x}, \mathbf{y}, \mathbf{z}, p) =
\begin{pmatrix}
x(t_1) \\ \vdots \\ x(t_M) \\
y(t_1) \\ \vdots \\ y(t_M) \\
z(t_1) \\ \vdots \\ z(t_M) \\
p
\end{pmatrix}.
\]

Therefore, to solve the equilibrium problem, we seek $(\mathbf{x}^*, \mathbf{y}^*, \mathbf{z}^*, p^*)$ satisfying the VI:
\begin{equation}
 \big\langle F(\mathbf{x}^*, \mathbf{y}^*, \mathbf{z}^*, p^*),\,
 (\mathbf{x}, \mathbf{y}, \mathbf{z}, p) - (\mathbf{x}^*, \mathbf{y}^*, \mathbf{z}^*, p^*) \big\rangle \geq 0, 
 \quad \forall (\mathbf{x}, \mathbf{y}, \mathbf{z}, p) \in K,
 \label{vi}
\end{equation}
with feasible set:
\[
K = \left\{ (\mathbf{x}, \mathbf{y}, \mathbf{z}, p) \,\middle|\, \mathbf{x}, \mathbf{y}, \mathbf{z} \in [0,1]^M,\ p \geq 0 \right\}.
\]

\noindent \textbf{Note} We do not explicitly add the constraint $x(t) + y(t) + z(t) \leq 1$ for each time period as in \autoref{early domain}. This is because the Logit model ensures that the mode choice probabilities $(\varphi_{\text{Car}}^t, \varphi_{\text{Bus}}^t, \varphi_{\text{DRAS}}^t)$ sum to 1 for any generalized cost vector. As the VI formulation drives the solution toward equilibrium where actual shares match these probabilities, the condition $x(t) + y(t) + z(t) = 1$ is satisfied endogenously in equilibrium.\\

\noindent \textbf{Proposition 1} The solution of 
A vector \((\mathbf{x}^*, \mathbf{y}^*, \mathbf{z}^*, p^*) \in K\) solves the VI within the domain $K$ if and only if:
\begin{equation}
(\mathbf{x}^*, \mathbf{y}^*, \mathbf{z}^*, p^*) = \Phi(\mathbf{C}(\mathbf{x}^*, \mathbf{y}^*, \mathbf{z}^*, p^*)).
\end{equation}

\noindent \textbf{Proof} 
Assume that \( (\mathbf{x}^*, \mathbf{y}^*, \mathbf{z}^*, p^*) \) is a solution to the VI problem, we have:
\begin{equation}
\langle F(\mathbf{x}^*, \mathbf{y}^*, \mathbf{z}^*, p^*), (\mathbf{x}, \mathbf{y}, \mathbf{z}, p) - (\mathbf{x}^*, \mathbf{y}^*, \mathbf{z}^*, p^*) \rangle \geq 0, \quad \forall (\mathbf{x}, \mathbf{y}, \mathbf{z}, p) \in K.
\end{equation}
Let \( (\mathbf{x}, \mathbf{y}, \mathbf{z}, p) = (\mathbf{x}^*, \mathbf{y}^*, \mathbf{z}^*, p^*) - \epsilon F(\mathbf{x}^*, \mathbf{y}^*, \mathbf{z}^*, p^*) \), where \( \epsilon > 0 \) is sufficiently small to ensure that \( (\mathbf{x}, \mathbf{y}, \mathbf{z}, p) \in K \). Substituting this into the VI condition yields:
\begin{equation}
\langle F(\mathbf{x}^*, \mathbf{y}^*, \mathbf{z}^*, p^*), -\epsilon F(\mathbf{x}^*, \mathbf{y}^*, \mathbf{z}^*, p^*) \rangle \geq 0.
\end{equation}
Simplifying this expression gives:
\begin{equation}
-\epsilon \|F(\mathbf{x}^*, \mathbf{y}^*, \mathbf{z}^*, p^*)\|^2 \geq 0.
\end{equation}
Since \( \|F(\cdot)\|^2 \geq 0 \), this inequality holds only if:
\begin{equation}
F(\mathbf{x}^*, \mathbf{y}^*, \mathbf{z}^*, p^*) = 0.
\end{equation}
By the definition of \( F(\cdot) \), this implies:
\begin{equation}
(\mathbf{x}^*, \mathbf{y}^*, \mathbf{z}^*, p^*) = \Phi(C(\mathbf{x}^*, \mathbf{y}^*, \mathbf{z}^*, p^*)).
\end{equation} 
Thus, the solution to the VI problem is a fixed point of the mapping \(\Phi(\mathbf{C}(\cdot))\).

\subsection{Discussion on the solution existence and uniqueness}

\subsubsection{Proof of existence}

\noindent \textbf{Proposition 2}: The proposed formulation has at least one solution.

\noindent \textbf{Proof} According to the \citep{facchinei2003finite}[2.2.3 Proposition.], let \( K \subseteq \mathbb{R}^n \) be closed, bounded, and convex, and let \( F : K \to \mathbb{R}^n \) be continuous, the $VI(K,F)$ has a solution.

In our formulation, the feasible region \( K \) is defined as:
\begin{equation}
K = \{(\mathbf{x}, \mathbf{y}, \mathbf{z}, p) \mid \mathbf{x}, \mathbf{y}, \mathbf{z} \in [0, 1]^M, \, p \geq 0\}.
\end{equation}

For the convenience of discussion of the mathematical properties, we define \(\Omega\) as the upper bound of \(p\). In practice, \(p\) is always bounded \citep{balzer2022modal}, making the domain $K$ bounded (\(\mathbf{x}, \mathbf{y}, \mathbf{z} \in [0, 1]\), \(0 \leq p \leq \Omega\)) and closed, as defined by linear inequalities and box constraints. In addition, the continuity of \( F(\mathbf{x}, \mathbf{y}, \mathbf{z}, p) \) ensured by \( C(\cdot) \) and \( \Phi(\cdot) \). Therefore, the existence of at least one solution to the VI problem in \autoref{vi} holds.

\subsubsection{Discussion on uniqueness}
In our context, the uniqueness of the VI solution is not guaranteed. The mapping \( F(x,y,z,p) \) is generally non-monotonic due to several structural features of the model. First, the endogenous credit price couples all time periods and modes through the market-clearing condition. Additionally, DRAS results in piecewise, non-monotonic dynamics in waiting times. As demand increases, riders typically experience decreasing waiting times up to the vehicle sufficient capacity threshold. Beyond this threshold, departing DRASs leave behind residual riders, which raises the waiting times for those remaining due to capacity constraints and spillover effects. These factors contribute to a Jacobian matrix whose symmetric part is not positive definite, thereby violating the standard strict monotonicity conditions that ensure uniqueness.

\section{Solving Algorithm}
A common and easy approach to solving VI problems is to reformulate the problem as an equivalent complementarity problem (CP) \citep{facchinei2003finite}, which can then be solved using the PATH solver \citep{dirkse1995path}. However, in our setting, the supply curves for both bus and DRAS modes involve a complex computation process among variables. This complication makes it challenging for the PATH solver to handle. Therefore, we develop a gradient-based optimization method tailored to our model structure. 

To solve for the mode choice equilibrium under this dynamic congestion setting, we go beyond the traditional Method of Successive Averages (MSA) \citep{lamotte2018morning,balzer2022modal}. Instead, we program the entire system, including travel time computation, congestion dynamics, generalized cost calculation, and discrete choice modeling, as a fully differentiable computation graph. This allows us to leverage automatic differentiation (autograd via  Pytorch) and apply gradient-based optimization directly. 

\subsection{Overall computation process}

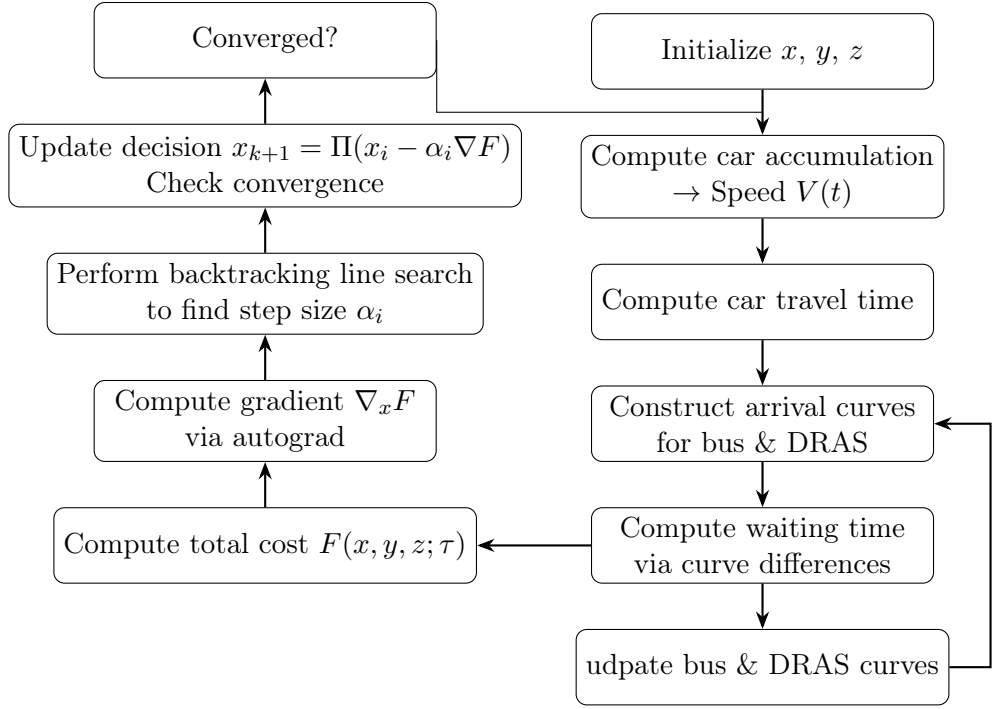
\begin{figure}[h]

\centering
\begin{tikzpicture}[
  node distance=0.6cm and 1.5cm,
  every node/.style={draw, rectangle, rounded corners, align=center, minimum width=4.5cm, minimum height=1cm},
  arrow/.style={-Stealth, thick}
]

\node (init) {Initialize $x$, $y$, $z$};
\node (speed) [below =of init] {Compute car accumulation \\ $\rightarrow$ Speed $V(t)$};
\node (car_tt) [below=of speed] {Compute car travel time };
\node (arrival) [below=of car_tt] {Construct arrival curves \\ for bus \& DRAS};
\node (wait) [below=of arrival] {Compute waiting time \\ via curve differences};
\node (updatecurve) [below=of wait] {udpate bus \& DRAS curves};
\node (cost) [left=of wait] {Compute total cost $F(x, y, z; \tau)$};
\node (grad) [above=of cost] {Compute gradient $\nabla_x F$ \\ via autograd};
\node (backtrack) [above=of grad] {Perform backtracking line search \\ to find step size $\alpha_i$};
\node (update) [above=of backtrack] {Update decision $x_{k+1} = \Pi(x_i - \alpha_i \nabla F)$ \\ Check convergence};
\node (check) [above=of update] {Converged?};
                                                
\draw[arrow] (init) -- (speed);
\draw[arrow] (speed) -- (car_tt);
\draw[arrow] (car_tt) -- (arrival);
\draw[arrow] (arrival) -- (wait);
\draw[arrow] (wait) -- (cost);
\draw[arrow] (cost) -- (grad);
\draw[arrow] (grad) -- (backtrack);
\draw[arrow] (backtrack) -- (update);
\draw[arrow] (update) -- (check);
\draw[arrow] (wait) -- (updatecurve);
\draw[arrow] (updatecurve) -- ++(3,0) |- (arrival);
\draw (check.east) |- ($(init.south)!0.5!(speed.north)$);

\end{tikzpicture}
\caption{Flowchart for Computation process}
\label{overalloop}
\end{figure}

To numerically obtain an equilibrium solution for the formulated VI problem, we incorporate several customized computational techniques tailored to our model structure. Specifically, the entire forward pass, including the travel time evaluation, disutility calculation, and mode choice computation, is executed iteratively (see \autoref{overalloop} for the full calculation loop).

\subsection{Projected Gradient with Backtracking}
Let $\lambda_0 \in K$ be an initial feasible decision vector, where 
\[
\lambda = (\mathbf{x}, \mathbf{y}, \mathbf{z}, p),
\]
and $k, \tau$ are fixed policy parameters. Here, 
$\mathbf{x}, \mathbf{y}, \mathbf{z} \in [0,1]^{MI}$ represent the time-dependent mode shares for Car, Bus, and DRAS across $M$ time intervals and $I$ OD groups, and $p \in \mathbb{R}_+$ is the credit price in the TCS.

At each iteration $i$, we compute the gradient $\nabla_\lambda F(\lambda_i; k, \tau)$, and update the candidate solution as
\[
\lambda_i^{\text{cand}} = \Pi_K\left( \lambda_i - \alpha_i \nabla_\lambda F(\lambda_i; k, \tau) \right),
\]
where $\Pi_K(\cdot)$ denotes projection onto the feasible set $K$, and $\alpha_i$ is the step size determined via backtracking line search.

The projection is applied elementwise:
\[
\lambda_i^{\text{cand}}[j] =
\begin{cases}
\min\big(\max(\lambda_i[j] - \alpha_i \nabla_j F, 0), 1\big), & j \leq 3MI, \\
\max(\lambda_i[j] - \alpha_i \nabla_j F, 0), & j = 3MI+1.
\end{cases}
\]

The backtracking condition requires sufficient descent:
\[
F(\lambda_i^{\text{cand}}; k, \tau) \leq F(\lambda_i; k, \tau) - c \cdot \alpha_i \| \nabla_\lambda F(\lambda_i; k, \tau) \|^2,
\]
with $c \in (0,1)$ (e.g., $10^{-4}$). 
If this condition fails, $\alpha_i \gets \beta \alpha_i$ (e.g., $\beta = 0.9$) until satisfied or the maximum retries is reached. 
When backtracking is unnecessary, $\alpha_i$ is increased to accelerate convergence. 
The algorithm terminates when the loss change falls below a tolerance $\varepsilon$.

\subsection{Selection on initial points}

In the above algorithm, we assume local convexity near the optimal solution. This assumption is reasonable in transportation systems, as policy changes typically result in gradual adjustments rather than abrupt shifts in traffic equilibrium. In practice, we implement this by first running the model without DRAS implementation and TCS to obtain a baseline equilibrium, and then sequentially introducing DRAS and the TCS policy. We use the baseline equilibrium as the initialization point for solving the new scenario, which brings two advantages: (i) from a behavioral and network dynamics perspective, gradual transitions make the baseline a stable and realistic starting point, especially with real dataset; and (ii) it narrows the search space, enhancing convergence speed and numerical stability. If the new policy induces substantial system changes, small perturbations or interpolations based on the baseline can be used to refine the initialization. 

\section{Numerical Study}
\begin{figure}[htbp]
\centering
\includegraphics[width=0.95\linewidth]{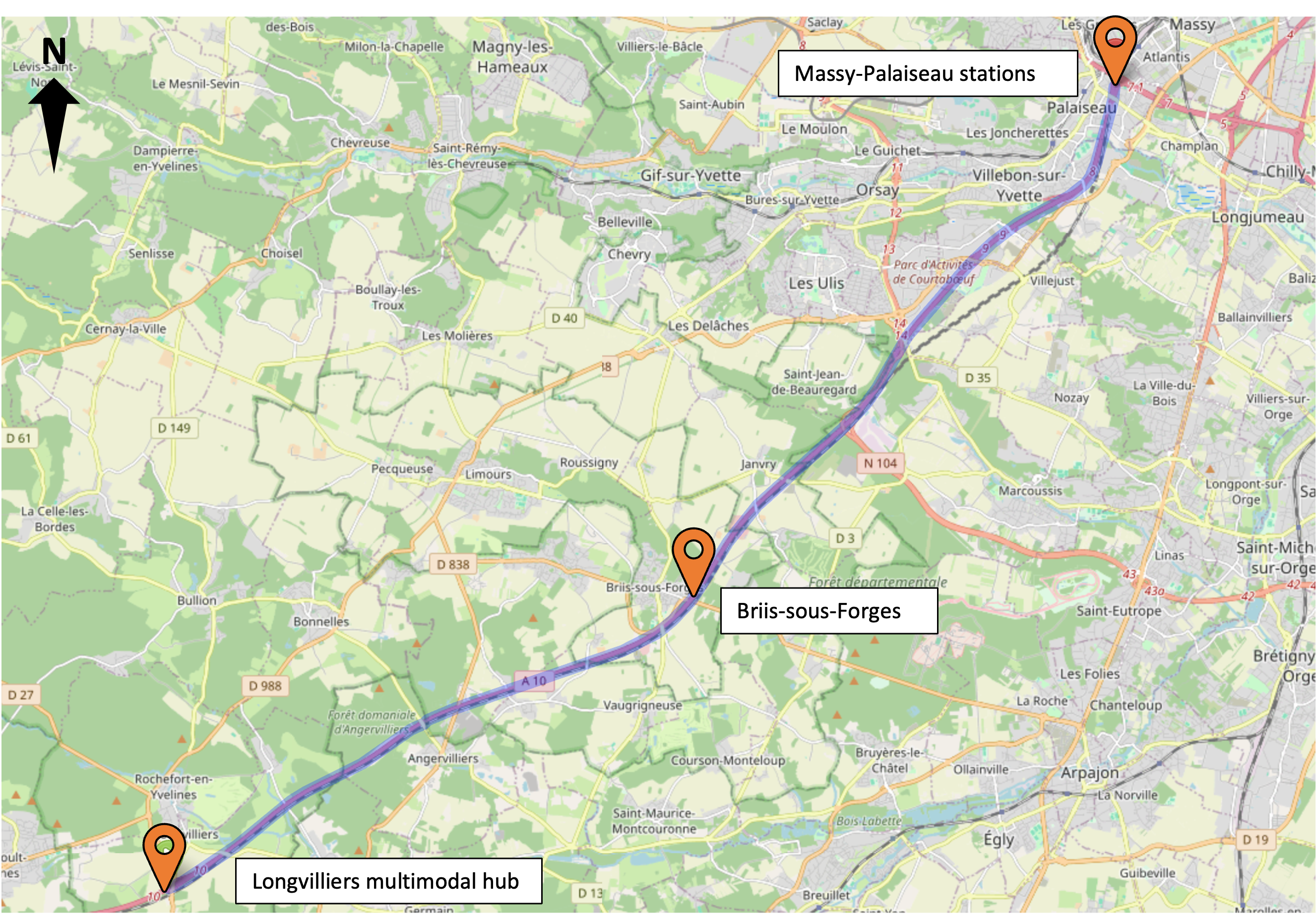}
\caption{The numerical study site. © OpenStreetMap contributors.}
\label{fig:study site}
\end{figure}

\begin{figure}[h]
    \centering
    \includegraphics[width=0.99\linewidth]{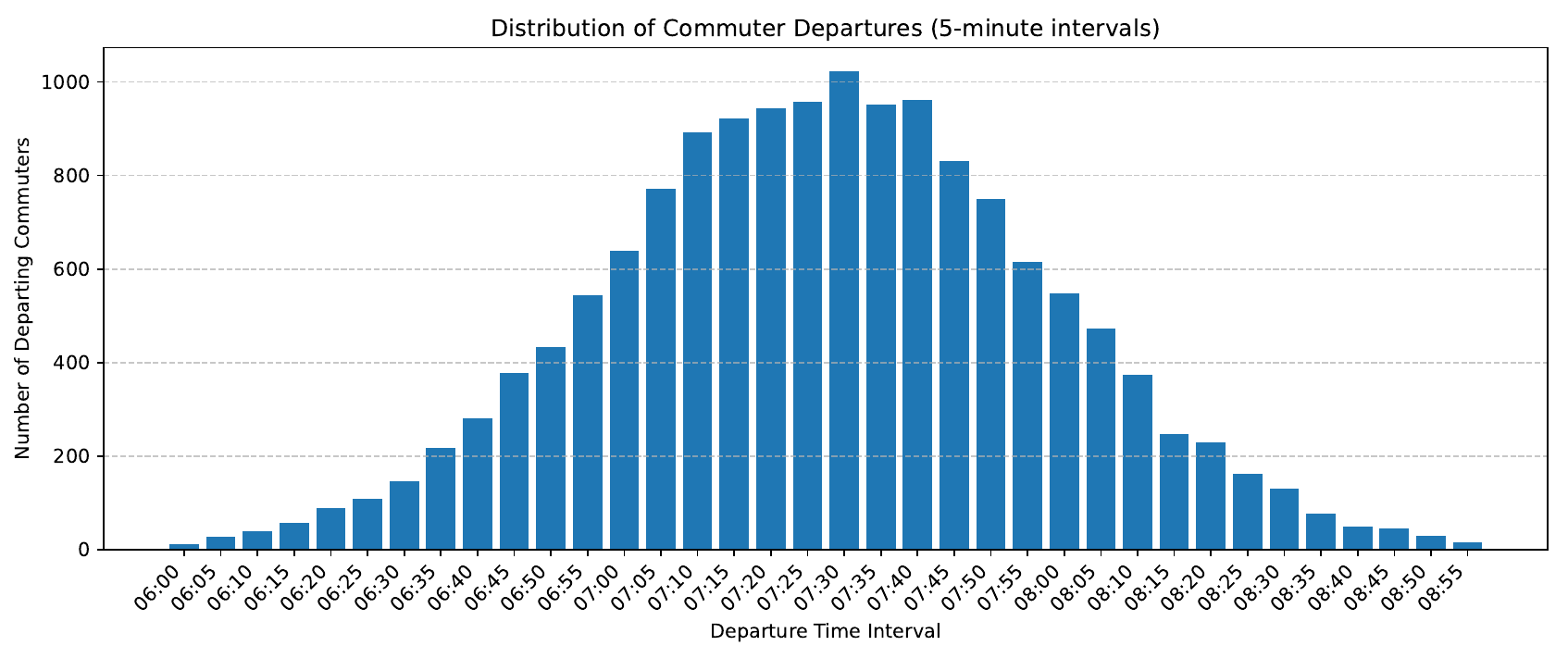}
    \caption{Distribution of Commuter Departures of the demand from Longvilliers to Massy-Palaiseau station }
    \label{fig:distribution}
\end{figure}

The numerical study focuses on a 28-kilometer segment of the A10 highway near Paris, France, starting from Longvilliers to Massy, passing through Briis-sous-Forges (12 km from Longvilliers to Briis, and 16 km from Briis to Massy), as shown in \autoref{fig:study site}. We assume conventional buses with 60-seat capacity depart every five minutes from the origin station, and five DRAS vehicles (each with 20-seat capacity) are dispatched with a minimum gap of four minutes and operate at a minimum one-minute headway. The sufficient factor $\omega = 0.8$ (i.e., departure while having 16 passengers), in-vehicle-VOT is set to $10.8$ \euro{}/hr according to the report \citep{fosgerau2007danish}, and waiting-VOT is set to $27.0$\euro{}/hr (1.5 to 2.5 times of in-vehicle VOT), according to the empirical studies \citet{wardman2004public}. Furthermore, we assume buses run at a maximum $90\; km/hr$ according to the normal setting of highway regulations in France, and DRASs run a bit slower at a maximum of $80\; km/h$ due to safety concerns. We simulate 15,000 travelers departing from Longvilliers to Massy-Palaiseau station during the morning peak. Departure times follow a normal distribution centered at 7:30 AM and are discretized into 36 five-minute intervals between 6:00 and 9:00 AM (see \autoref{fig:distribution}).
\label{num_setting}

Similarly, we simulate 3,000 travelers commuting from Briis-sous-Forges to Massy-Palaiseau station during the same morning peak period, with departure times drawn from the same normal distribution and discretized similarly. These demand estimates are derived from empirical count data. The travel demand from Longvilliers to Briis is much smaller; we set it to 200 in the simulation, with departure times generated in the same manner as above.

\begin{figure}{}
    \centering
    \includegraphics[width=0.9\linewidth]{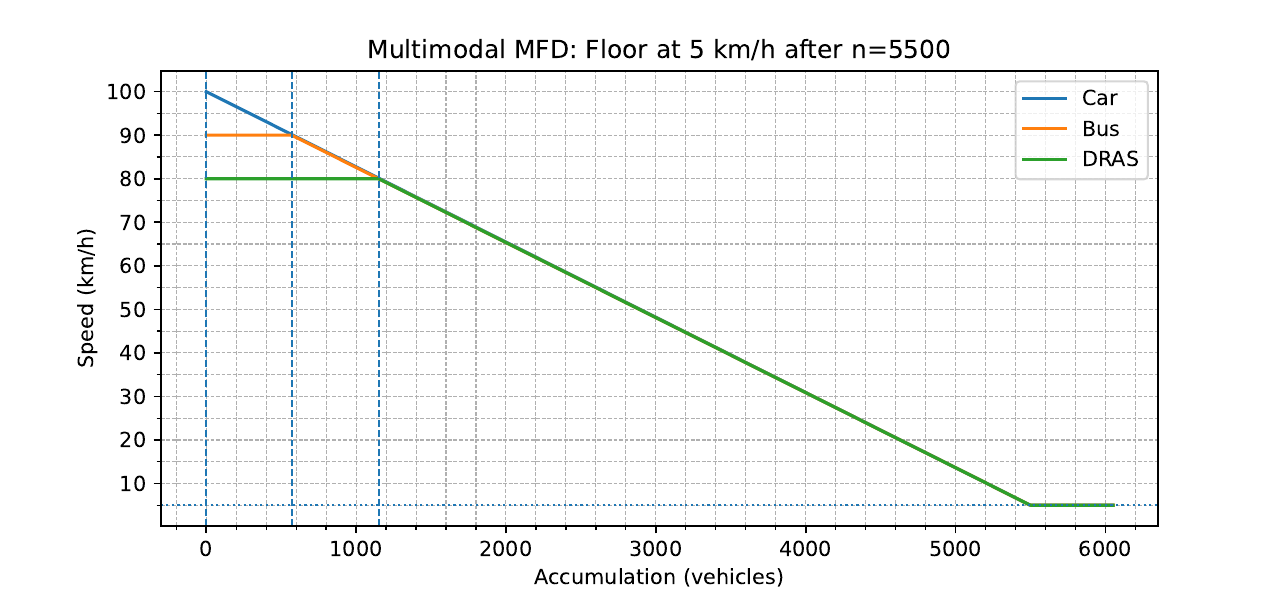}
    \caption{Multimodal accumulation–speed curves with free-flow caps}
    \label{fmfd}
\end{figure}

To capture the effects of congestion dynamics, we use a piecewise multimodal MFD speed-accumulation function,

$$
v(n) = v_{\text{max}} \left( 1 - \frac{n}{n_{\text{max}}} \right),
$$

where $v_{\text{max}}$ denotes the free-flow speed cap of each mode (e.g., $100$ km/h for cars, $90$ km/h for buses, and $80$ km/h for DRAS), and $n_{\text{max}} = 5500$ vehicles is the maximum accumulation. To reflect highway operations and avoid unrealistic gridlock, we impose a minimum speed floor of $5\ \text{km/h}$. (see \autoref{fmfd} as a reference)

The parameter settings of the numerical study are summarized in Table \ref{tab:parameters}.
\begin{table}[h]
\centering
\caption{Fixed model parameters}
\begin{tabular}{lll}
\hline
\textbf{Parameter} & \textbf{Value} & \textbf{Description} \\
\hline
$L$ & 28,000 m & Road length \\
$v^{\max}$ & 100 km/h & Maximum speed \\
$v^{\max}_{bus}$ & 90 km/h & Maximum speed for buses \\
$v^{\max}_{DRAS}$ & 80 km/h & Maximum speed for DRASs\\
$v^{\min}$ & 5 km/h & Minimum speed (to avoid gridlock) \\
$n^{\max}$ & 5,500 veh & Maximum vehicle accumulation \\
$t_{\text{segment}}$ & 300 s & Simulation time step \\
$\alpha$ & 10.5 \euro{}/h & In-vehicle value of time \\
$\alpha'$ & 26.5 \euro{}/h & Value of time for waiting \\
$\theta$ & 0.1 & Utility scale parameter \\
Bus capacity & 60 pax & Capacity per bus \\
Bus interval & 10 min & Bus departure interval \\
DRAS capacity & 20 pax & Capacity per DRAS vehicle \\
Minimum departure headway for DRAS & 1 min & The minimum time interval between \\ consecutive DRAS vehicle departures.\\

\hline
\end{tabular}
\label{tab:parameters}
\end{table}

To examine the core behavioral and operational mechanisms, we first conduct a parametric analysis in a single-OD setting before considering the multi OD setting for an optimum balanced solution. In the single-OD study, we systematically vary key parameters, such as the DRAS fleet size, the credit allocation ratio $k/\tau$, and the minimum occupancy threshold for dispatching DRAS, to evaluate their impact on system performance. Performance metrics include total travel time (including and excluding waiting time), total waiting time, and equilibrium credit price. This analysis helps demonstrate fundamental system dynamics and reveal economic and behavioral insights under different policy or operational interventions. Later, we extend the model to a multi-OD network, where mode choices and congestion effects interact across different OD pairs. In this more realistic setting, we formulate and solve a bi-level optimization problem that jointly determines the optimal fleet size and credit scheme.

\subsection{Results on consideration of operational constraints}
Before presenting the parametric analysis, we first numerically demonstrate the role of operational considerations in shaping mode choice outcomes. We compare two scenarios: the first scenario ignores the operational and queuing mechanisms introduced in \autoref{fig:structure}, while the second explicitly incorporates them. This comparison provides a direct benchmark, showing how the absence or presence of these modules fundamentally alters the resulting system dynamics and, consequently, the interpretation of modal attractiveness.

\begin{figure}[htbp]
    \centering
    \includegraphics[width=1\linewidth]{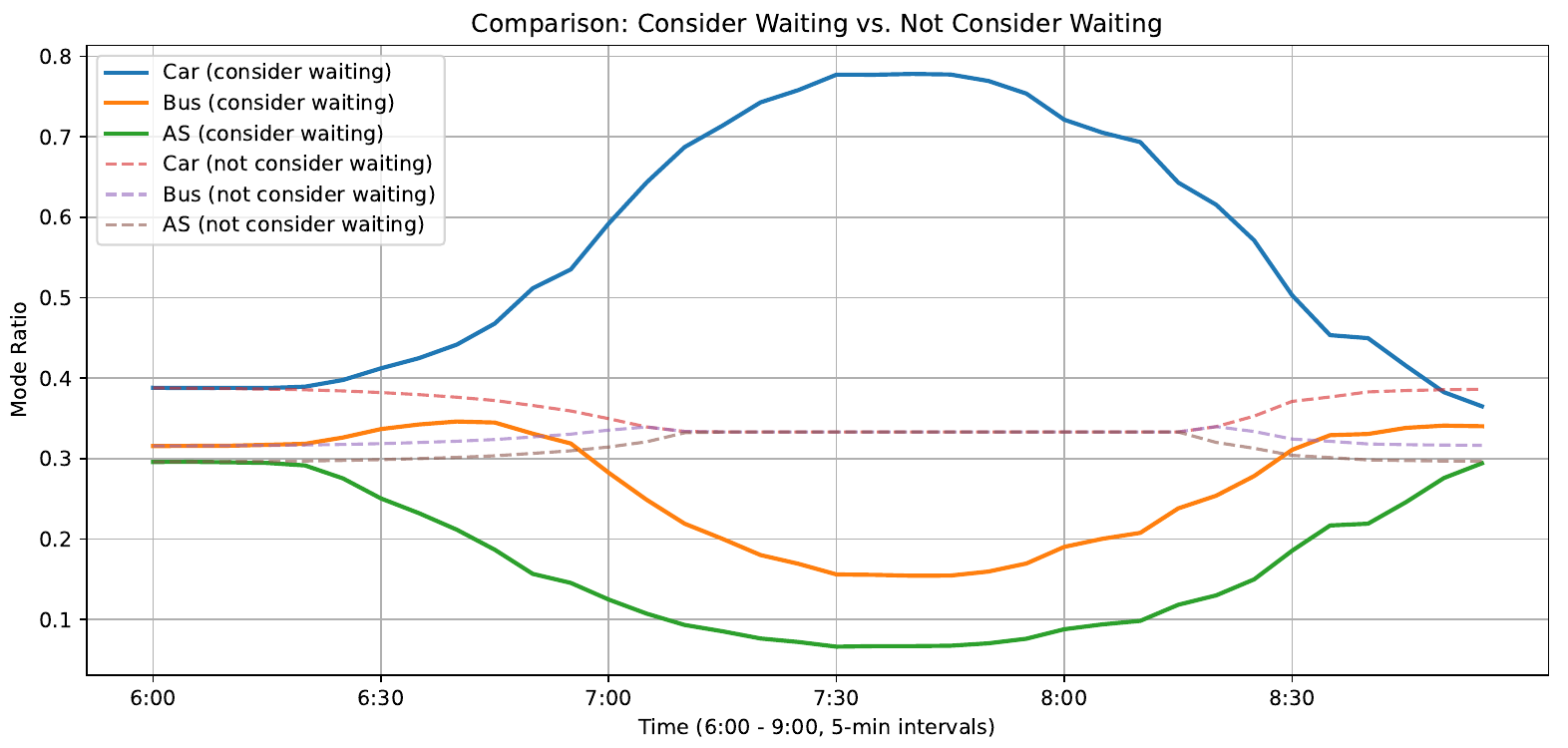}
    \caption{Comparison of mode shares under scenarios with and without operational features}
    \label{fig:operation ratios}
\end{figure}

As shown in \autoref{fig:operation ratios}, for the baseline scenario (i.e., without consideration of operational features), the attractiveness of different mode is distinguishable before congestion emerges, which is driven by the speed-density curve of the multimodal MFD in \autoref{fmfd}. As shown in the plot, early in the morning with no congestion, the car option has the highest mode share due to its highest free flow speed, followed by bus's and DRAS's. As the congestion emerges, the mode share starts to converge. Because all three modes experience the same congestion speed, the time-based cost essentially overlaps. Once the accumulation of cars decreases and congestion falls, their speeds rise to free-flow speeds again, then the distinguished attractiveness appears again. As a result, the mode share ratios diverge at the beginning and end of the study period, but converge toward an equal split of approximately 33.3\% in the congested period, where demand is distributed across modes with no differentiation. Notably, because buses and DRASs are modeled without waiting times, their generalized costs during congestion are equal to the car's generalized cost (e.g., solely driven by travel time), which is significantly underestimated, while the car’s mode ratio is consequently biased downward.

In contrast, incorporating operational features generates markedly different mode share dynamics. Station-level waiting times, caused by service capacity and dispatch rules, introduce additional time-based costs that differentiate the generalized cost of bus's and DRAS's from car’s, especially during congestion. As demand grows, limited fleet size and scheduling rules cause queuing delays to raise. DRAS, with a more limited fleet, shows an earlier decline in modal share as congestion emerges. Bus initially absorbs some displaced travel demand, but as congestion intensifies and station queues lengthen, its share also decreases. Meanwhile, car becomes relatively more attractive despite roadway congestion, since it avoids waiting time on boarding their private vehicles. This operationally enriched representation reflects a more realistic allocation of passengers across modes, where operational frictions play as important as roadway congestion in shaping mode choice.

These observations underscore the importance of incorporating operational features when analyzing multimodal systems. Building on this distinction, we next turn to a parametric analysis in a single-OD setting to disentangle the specific effects of credit allocation, fleet size, and dispatch thresholds on system performance.

\subsection{Parametric Analysis on Single OD scenario}
We now consider a single OD pair with no intermediate access points within the morning peak hours, such that there is no demand traveling from Briis-sous-Forges to Massy-Palaiseau station, but 15,000 travel demand from Longvilliers to Massy. We are testing the implementation of two DRASs for this single OD scenario, excluding the study on the number of DRAS implementations.

\subsection{Results on consideration of TCS}
We numerically demonstrate the role of TCS in shaping mode choice outcomes. We compare two scenarios: the first scenario considers the operational and queuing mechanisms, but does not consider the TCS implementation in the system, while the second scenario considers the TCS implementation. This side-by-side comparison provides a direct benchmark, showing how the absence or presence of TCS fundamentally alters the resulting system mode shares.

We adopt the same experimental setting as in \autoref{fig:operation ratios}, and further use the scenario that considers operational constraints from \autoref{fig:operation ratios} as the baseline. In contrast, with the TCS scenario, each traveler is allocated $5$ credits (i.e., $k=5$) and asks for $8$ credits if they drive their private cars ($\tau = 8$), so travelers who drive need to purchase extra $3$ credits, and passengers who take public or shared ride could sell their $5$ credits for monetary compensation. The result shows that by imposing tradable credits on car users, TCS raises the generalized cost of driving, effectively discouraging car usage during peak periods (i.e., from 6 am to 9 am) and redirecting demand to buses and DRAS. However, the most significant changes occur before the onset of heavy congestion (before 7:00) and after congestion dissipates (after 8:30). The reasons for this is that, although TCS increases the generalized cost of driving private vehicles, diverting a large share of demand to public and shared modes, but the induced waiting time disutility, due to operational constraints, outweigh the monetary surcharge imposed by TCS. Thus, when congestion is not severe and public/shared modes can accommodate the additional demand, the rise in TCS costs makes mode switching away from private cars.

\begin{figure}[htbp]
    \centering
    \includegraphics[width=1\linewidth]{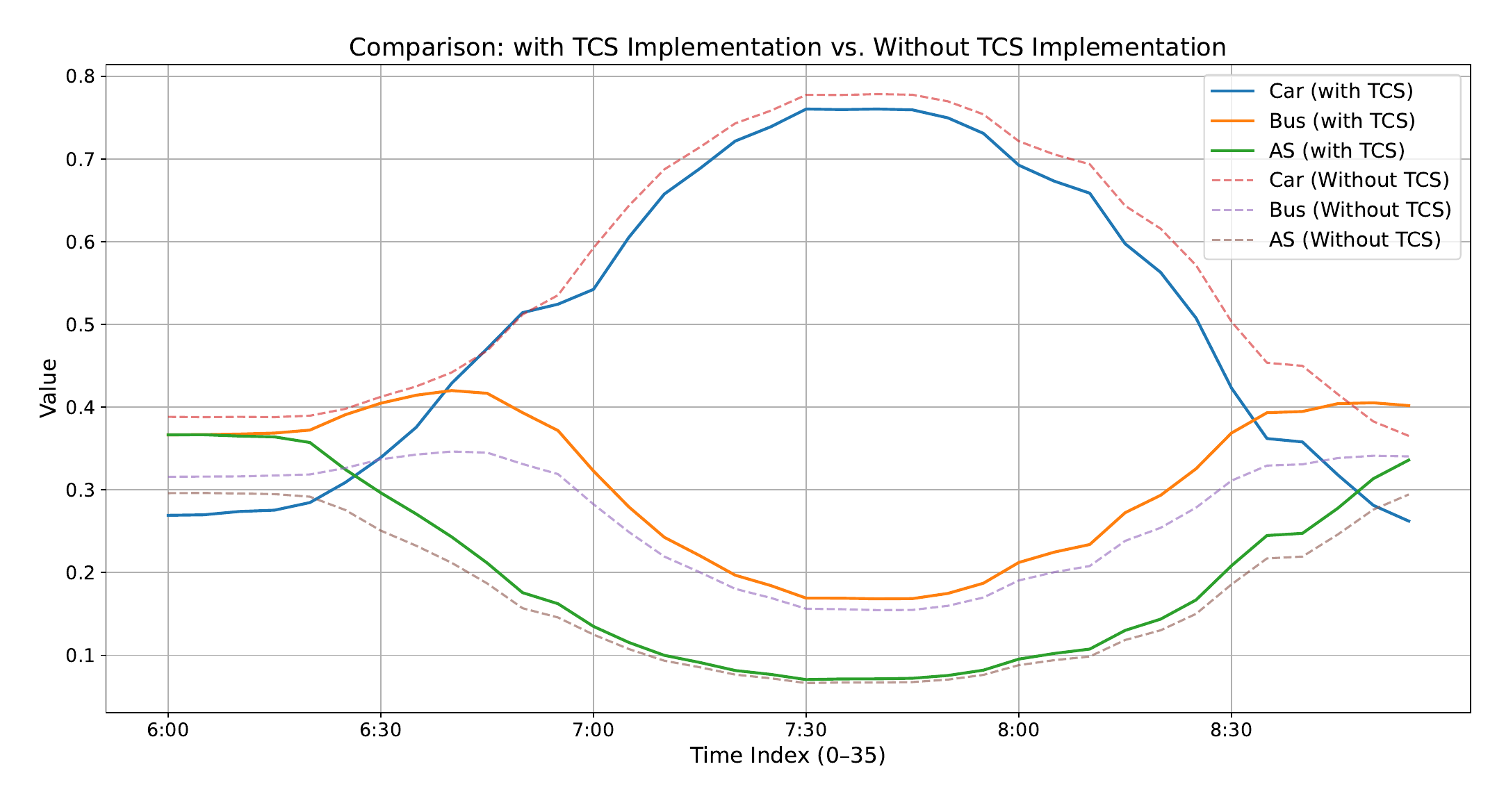}
    \caption{Comparison of mode shares under scenarios with and without TCS implementation}
    \label{fig:ratios}
\end{figure}

The TCS raises the generalized cost of driving and shifts demand from private cars toward buses and DRAS. To better observe and quantify these mechanisms, we vary the credit endowment $k$ and the per-trip requirement $\tau$ and examine different performances in the next section.

\subsubsection{Parametric Study of the TCS Scheme}
We first define a variable, maximum allowable driving rate, $d_{max}$ as the ratio of $k$ over $\tau$, indicating the largest allowable share of drivers for the study period. To see this more clearly: suppose there are $N$ travelers and each receives $k$ credits, while a single car trip requires $\tau$ credits. The total endowment is $Nk$, so at most $Nk/\tau$ trips can be undertaken. Normalizing by $N$ yields an upper bound on the driving fraction, $k/\tau$.

We start by varying the initial allocated number of credits $k$ from $50$ to $58$ in increments of $2$. For each $k$, we vary $\tau$ from $64$ to $72$ in increments of 2. Therefore, we have the following grid table \autoref{ratiotable} showing the $d_{max}$ for different combinations of $k$ and $tau$. For each combination, we compute performance metrics, including total travel time, total waiting time, and credit price at equilibrium status. 
\begin{table}[h!]
\centering
\caption{$k/\tau$ as percentages}
\begin{tabular}{c|ccccc}
\hline
$\tau \backslash k$ & 50 & 52 & 54 & 56 & 58 \\
\hline
72 & 69.4\% & 72.2\% & 75.0\% & 77.8\% & 80.6\% \\
70 & 71.4\% & 74.3\% & 77.1\% & 80.0\% & 82.9\% \\
68 & 73.5\% & 76.5\% & 79.4\% & 82.4\% & 85.3\% \\
66 & 75.8\% & 78.8\% & 81.8\% & 84.8\% & 87.9\% \\
64 & 78.1\% & 81.2\% & 84.4\% & 87.5\% & 90.6\% \\
\hline
\end{tabular}
\label{ratiotable}
\end{table}

\begin{figure}
    \centering
    \includegraphics[width=1\linewidth]{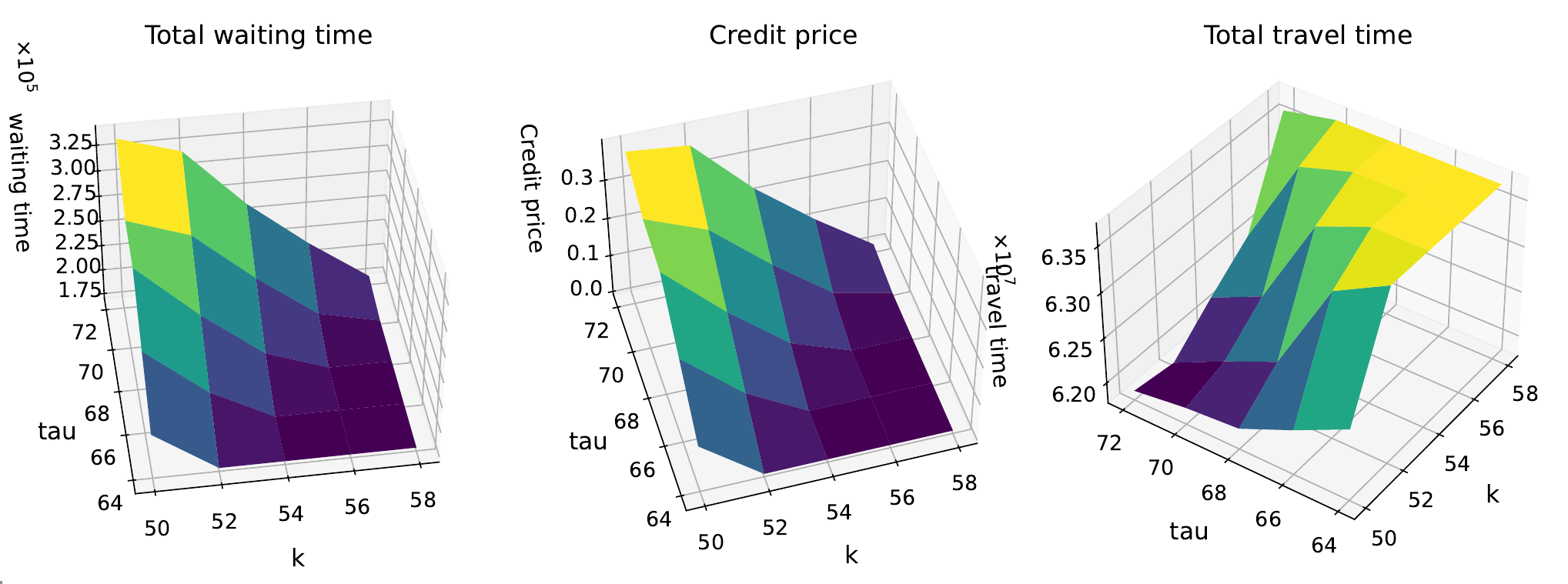}
    \caption{The 3D Surface plots of different performance metrics}
    \label{fig:katau}
\end{figure}

As shown in \autoref{fig:katau}, we present the surfaces for our four performance metrics over the five-by-five grid of $k\in{50,52,54,56,58}$ and $\tau\in{64,66,68,70,72}$. For clarity of presentation, we do not overlay $d_{\max}$ on the figures; the corresponding $d_{\max}$ can be found in \autoref{ratiotable}. We summarize the main findings below.

First, the credit price surface shows a monotonic trend with respect to the tightness of the tradable credit constraint. Specifically, prices rise systematically as the ratio $d_{max}$ decreases (i.e., as the policy becomes more restrictive). When the $d_{max}$ is loose (i.e., initial credit endowment $k$ is relatively high or the required credits per trip $\tau$ is low), the constraint remains slack and the credit price remains at zero. For instance, the driving ratio without policies is $83\%$, and when $d_{max}$ = $87.5\%$ and $90.6\%$ (i.e., $\tau = 64$, and $k = 56 $ or $54$), the price is exactly zero, indicating no scarcity. However, as the $d_{max}$ decreases, the market becomes constrained and prices begin to rise. At $d_{max}$ = $78.1\%$ (i.e., $k = 50$ and $\tau = 64$), the credit price is \euro{}0.11 (the out-of-pocket expense is \euro{} 1.32 per trip), and it increases to a peak of about \euro{}0.36 (the out-of-pocket expense is \euro{} 4.32 per trip) at $d_{max}$ = $69.4\%$ (i.e., $k = 50$ and $\tau = 72$). This pattern reflects the endogenous response of the credit market to the tightening supply–demand balance.

Second, the total waiting time increases monotonically as the $d_{max}$ decreases. As more travelers are expected to be diverted from private car usage due to tighter TCS constraints, pressure on shared modes grows, resulting in longer boarding queues. For example, total waiting time increases from roughly 17.9 thousand seconds when $d_{max}$ =  $87.5\%$ (i.e.,$\tau=64,k=56$) to over 31.0 thousand seconds at $d_{max}$ = $69.4\%$ (i.e., $k = 50$ and $\tau = 72$). 

Finally, the total travel time, accounting for both in-vehicle and waiting times, exhibits a decreasing trend as the $d_{max}$ decreases, where the TCS policy becomes more stringent. Without TCS bounding, the system records the highest total travel time, exceeding $63$ million seconds. In contrast, under the most restrictive policy condition where $d_{max}$ = $69.4\%$ (i.e., $k = 50$ and $\tau = 72$), the total travel time drops to approximately $61.5$ million seconds. 


This pattern suggests that the benefits of traffic congestion relieve (shorter in-vehicle time) outweigh the costs of increased waiting under tighter TCS constraints. Unlike what might be expected in systems with strong transit bottlenecks, the rise in waiting time here is not large enough to reverse the overall gains in travel efficiency. Hence, from a system-level perspective, stronger constraints consistently lead to better outcomes in terms of total travel time. 

\subsubsection{Parametric Study of DRAS Implementation}
We test the impact of number of DRAS serving in the corridor. We vary the maximum allowable driving rate $d_{max}$ from $70\%$ to$78\%$ in increments of $2\%$. For each combination, we compute performance metrics, including total travel time, total waiting time, and credit price at equilibrium status. 
\begin{figure}[!htbp]
    \centering
    \includegraphics[width=1\linewidth]{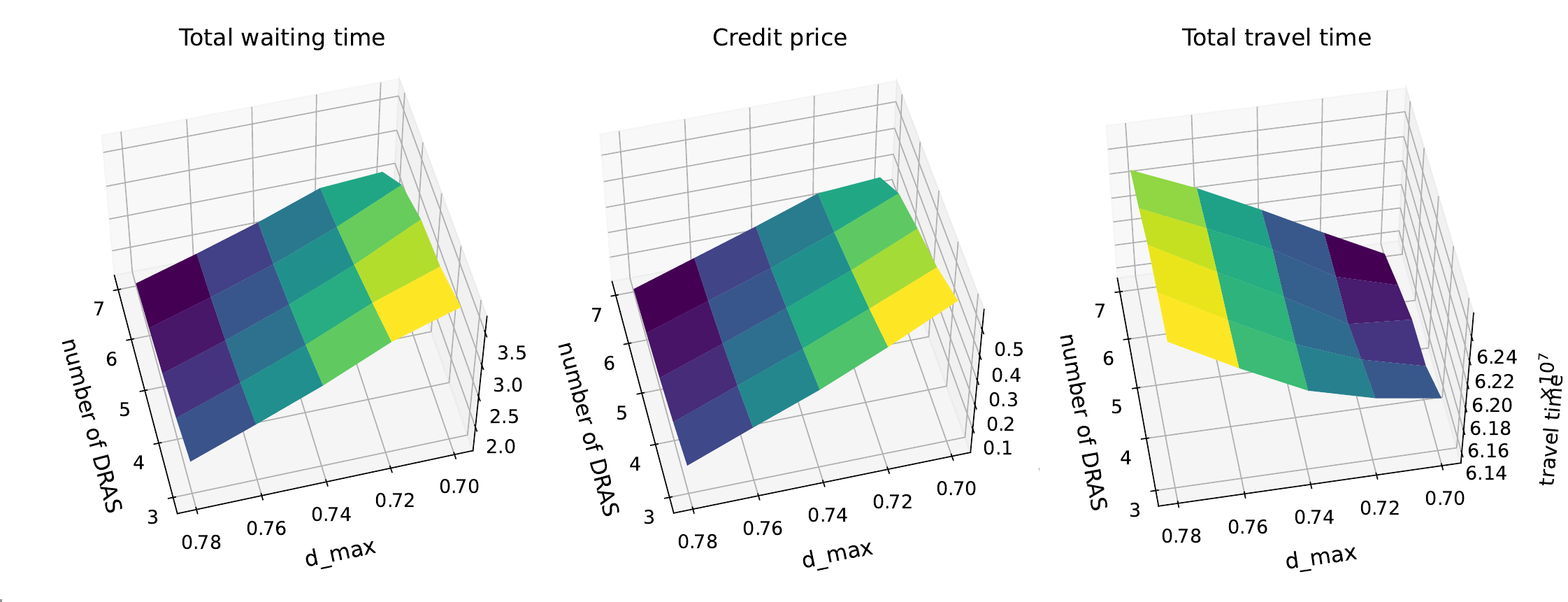}
    \caption{3D surface plots of different performance metrics under different number of DRAS, given fixed $k$ and $\tau$}
    \label{fig:asvariation}
\end{figure}

As shown in \autoref{fig:asvariation}, the solution surface generally shows a monotonic pattern; however, it shows fluctuations that are not consistent. This inconsistency may be due to the piecewise and non-monotonic dynamics of waiting times in the DRAS.
Furthermore, larger fleets consistently improve performance by lowering total travel time and passenger waiting. For instance, under a driving allowable rate of $78\%$, raising the number of DRAS vehicles from 3 to 7 cuts total travel time by nearly $2$ million seconds. Waiting times exhibit a similar pattern: additional vehicles sharply reduce delays when the fleet grows from 3 to 7, but under tight credit constraints, waiting remains considerable even with 7 vehicles, underscoring the pressure that stricter policies place on shared modes. Credit prices also decline with fleet expansion, reflecting the system’s improved ability to absorb travelers priced out of car use. This effect is strongest at moderate $\tau$ levels, though prices stay high under very restrictive policies, suggesting that DRAS supply alone cannot fully offset regulatory pressure.

\section{Multi-OD Bi-Level Optimization}
Building on the insights from the single-OD equilibrium analysis, we extend the model to a more realistic setting with multiple OD pairs. For the multiple OD scenario, we consider three OD pairs: from Longvilliers to Massy, from Longvilliers to Briis, and Briis to Massy, and follow the setting in \autoref{num_setting}. In this extended setting, we develop a bi-level optimization framework that jointly determines the deployment of DRAS fleets and the design of the TCS, while explicitly accounting for travelers’ behavioral responses as characterized by the equilibrium model (see \autoref{equlibrium}). In addition, we account for through-traffic that traverses the corridor without participating in either the TCS or the mode choice process. Although these vehicles are not directly influenced by the policy instruments under consideration, their presence affects congestion dynamics and contributes to a more realistic and complex representation of system performance.

The single OD analysis section highlighted how both TCS and DRAS interventions influence multimodal travel behavior and overall system performance. The results reveal two important insights. First, given the fixed number of credits $k$, increasing the required number of credits effectively discourages private car use by raising its generalized cost under TCS. However, when public and shared transit capacity is limited, this displaced demand can overwhelm available services capacity, leading to extended waiting times, and ultimately, a deterioration in the travel experience. Second, increasing the DRAS fleet size helps alleviate this capacity pressure by offering a flexible and responsive transport alternative. This reduces station-level waiting times and improves user experience. However, expanding the DRAS fleet entails substantial capital investment. Thus, while a greater DRAS supply improves network performance, it also imposes significant financial costs on planners or operators.

To summarize, these two findings reveal two fundamental trade-offs that would be considered in bi-level optimization design:
\begin{enumerate}
    \item The trade-off between reducing private car usage through stricter credit policies and the risk of overwhelming limited transit capacity.
    \item The trade-off between reducing waiting times through increased DRAS supply and the associated capital investment costs.
\end{enumerate}    

\subsection{Bi-level formulation}
To optimize system performance and balance the aforementioned trade-offs, we initially considered a weighted sum of two components: total travel time and the deployment cost of DRAS vehicles. Furthermore, we additionally incorporate vehicle-kilometers traveled (VKT) into the objective function and take as a proxy for emissions and energy use. In this way, the objective jointly captures temporal performance, fiscal expenditure, and environmental externalities. At the lower level, travelers respond to the given operational settings and policy design by choosing travel modes that minimize their individual disutility, leading to a stochastic user equilibrium.

This behavioral response is formalized through a complementarity-based formulation, where mode choices arise endogenously as equilibrium outcomes under generalized travel costs. As a result, the overall system can be reduced as a Mathematical Program with Equilibrium Constraints (MPEC), in which the upper-level planner’s optimization problem is subject to the equilibrium conditions of travelers at the lower level. The decision variables are $\theta = \{k, \tau, \xi\}$, where $k$ is the initial credit allocation, $\tau$ is the required number of credits for driving private vehicles, and $\xi$ represents the number of DRAS for operation. The variables $x(t)$, $y(t)$, and $z(t)$, previously defined in the equilibrium formulation, denote the number of users choosing car, bus, and DRAS modes at time $t$, respectively. We use $\alpha$, the in-vehicle VOT, as the conversion factor for the travel time to monetary expense. Besides, we use $\beta_{em}$ as the conversion factor. This factor converts vehicles’CO$_2$ emission rate into its equivalent carbon offset cost per meter. Given an emission rate of 120 grams per kilometer per vehicle and an offset price of \euro{} 100 per ton of CO$_2$, the result is 1.2 cents \euro{} per kilometer (or 0.000012 \euro{} per meter). This value represents how much it would cost to fully offset the carbon emissions for each meter the vehicle travels. In addition, the current framework does not consider the price elasticity of demand, which means that travelers are willing to accept any price, no matter how high it is. This oversight can produce unrealistic model outcomes, potentially leading to distorted policies and poor decision-making. To address this issue, we incorporate credit price as a soft bound constraint in the objective function. This strategy makes excessively high prices more burdensome, encouraging a balanced trade-off between supply settings, credit quota design, DRAS deployment cost, and price levels. We denote $\gamma_{tt}$, $\gamma_{em}$, $\gamma_{as}$ and $\gamma_{cp}$ as the weights assigned to different components of the objective function: travel time ($\gamma_{tt}$), distance traveled ($\gamma_{em}$), fleet cost ($\gamma_{as}$) and credit price ($\gamma_{cp}$). Thus, we have the reduced MPEC formulation as follows:



\begin{align}
\min_{\theta = \{k,\tau,\xi\}} \quad 
& \underbrace{\gamma_{tt} \cdot\alpha\cdot \sum_{i,t,\mu} T_{i}^{\mu}(t)}_{\text{Total travel time}}
+ \underbrace{\gamma_{em} \cdot\beta_{em} \cdot \sum_{t} n^{\text{car}}(t) \cdot l^{\text{car}}}_{\text{Emission}}
+ \underbrace{\gamma_{as} \cdot \text{FleetCost}_{\text{DRAS}}(\theta)}_{\text{Fleet investment cost}} \label{eq:obj} + \underbrace{\gamma_{cp} \cdot p}_{\text{Credit price}}\\
\text{subject to:} \quad 
& 0 \leq x(t) \perp x(t) - \varphi_{\text{Car}}^t(\mathbf{C}) \geq 0,\quad \forall t = 1, \dots, M \label{eq:car_equilibrium} \\
& 0 \leq y(t) \perp y(t) - \varphi_{\text{Bus}}^t(\mathbf{C}) \geq 0,\quad \forall t = 1, \dots, M \label{eq:bus_equilibrium} \\
& 0 \leq z(t) \perp z(t) - \varphi_{\text{DRAS}}^t(\mathbf{C}) \geq 0,\quad \forall t = 1, \dots, M \label{eq:as_equilibrium} \\
& 0 \leq p \perp (k - \tau) \cdot \sum_{t=1}^M x(t) \geq 0 \label{eq:price_complementarity} \\
& \xi b \leq B \label{eq:budget} \\
& x(t)\geq 0, y(t) \geq 0, z(t) \geq 0, \xi \geq 0,\quad \forall t \label{eq:nonnegativity}
\end{align}

The mode choice behavior is modeled using complementarity conditions \autoref{eq:car_equilibrium}, \autoref{eq:bus_equilibrium}, and \autoref{eq:as_equilibrium} ensure no user can unilaterally switch modes to reduce their travel cost $\varphi^\mu_t(\mathbf{C})$ at equilibrium conditions. \autoref{eq:price_complementarity} demonstrates the MCC condition that all the initial allocated credits are used at equilibria $k - \tau \sum_t x(t)$. \autoref{eq:budget} demonstrate the number of DRAS $\xi$ times the unit cost of DRAS deployment $b$ should be bounded by $B$. Finally, non-negativity constraints ensure all flow and investment variables are physically meaningful.
\subsection{Solving algorithm for the bi-level optimization}
Solving the proposed bi-level model is particularly challenging due to the non-convexity of the equilibrium constraints. The addition of upper-level decision variables further increase the complexity, making it difficult to identify global optima. Existing optimization techniques often offer limited effectiveness for solving large-scale, non-convex bi-level problems. However, in practical applications, the domain of decision variables, such as fleet size and credit policy parameters, is naturally bounded by physical, budgetary, and regulatory constraints. This bounded and discrete nature of the variable domain allows for a more tractable search process, enabling the identification of high-quality solutions without the need for computationally intensive global optimization methods.

In practice, the number of initial credits allocated $k$, the number of credit requirements per car trip $\tau$, and the fleet size of DRAS is neither continuous nor unbounded. These parameters are typically chosen from a small set of feasible values based on operational feasibility and policy acceptability. For instance, the maximum allowable driving ratio ($d_{max}$) can't be too low considering the acceptance of TCS and the driving rights for the majority of travelers. To ensure these, the maximum allowable driving ratio cannot fall below a certain threshold (e.g., 0.6 or 0.5), making the feasible set of $d_{max}$ values limited and interpretable. Similarly, the number of DRAS vehicles cannot be scaled arbitrarily due to physical constraints, such as budget limits and minimum headway requirements. On a shared urban corridor, we set the headway as one minutes. These natural restrictions sharply reduce the dimensionality and granularity of the decision space.
 
Given this context, we adopt a discrete grid search approach to explore all potential values of $\tau,k,\xi$. For each configuration, we solve the lower-level equilibrium using the algorithm developed earlier, and compare the resulting system performance in terms of travel time, emissions, and operational costs. By comparing all local optima within the search domain, we identify the global optimum. Specifically, we perform a discrete grid search over $9$ values of the $d_{max}$, and $7$ values of DRAS fleet sizes from $4$ to $10$ vehicles. We allocate $50$ credits for each traveler and charge $60$ to $68$ credits with a $ 1$-increment for driving their private cars separately in different scenarios. Therefore, the maximum allowed driving ratio for a traveler is approximately $83\%,82\%,81\%,79\%,78\%,77\%,76\%,75\%,74\%$, respectively.

The weights assigned to different components of the objective function, travel time $\gamma_{tt}$, distance traveled $\gamma_{em}$, fleet cost $\gamma_{as}$, and credit price $\gamma_{cp}$, reflect the priorities of a central controller and are not intended to represent universal valuations. In practice, these parameters would depend on the policymaker's emphasis on user experience, environmental externalities, infrastructure investment, and users' cost. Our primary goal, however, is to ensure a meaningful comparison across policy scenarios. Therefore, we adopt a parameter setting where $\gamma_{tt}$ and $\gamma_{em}$, are both set to 1. These values are not intended to reflect empirical valuations of time or distance, but rather to normalize the magnitude of the travel time, distance, and monetary cost, bringing them onto a scale comparable to the capital cost term. The cost of deploying a DRAS is assumed to be \euro{}1 million per vehicle, including acquisition, maintenance, and insurance, and is amortized over a 10-year lifespan, resulting in a daily cost of approximately \euro{}274 per vehicle. We further set the $\gamma_{as}$ and $\gamma_{cp}$ as 0.5 and 0.05.

\subsection{Numerical results}
\subsubsection*{Results without considering credit price}
\begin{figure}[htbp]
    \centering
    \includegraphics[width=0.95\linewidth]{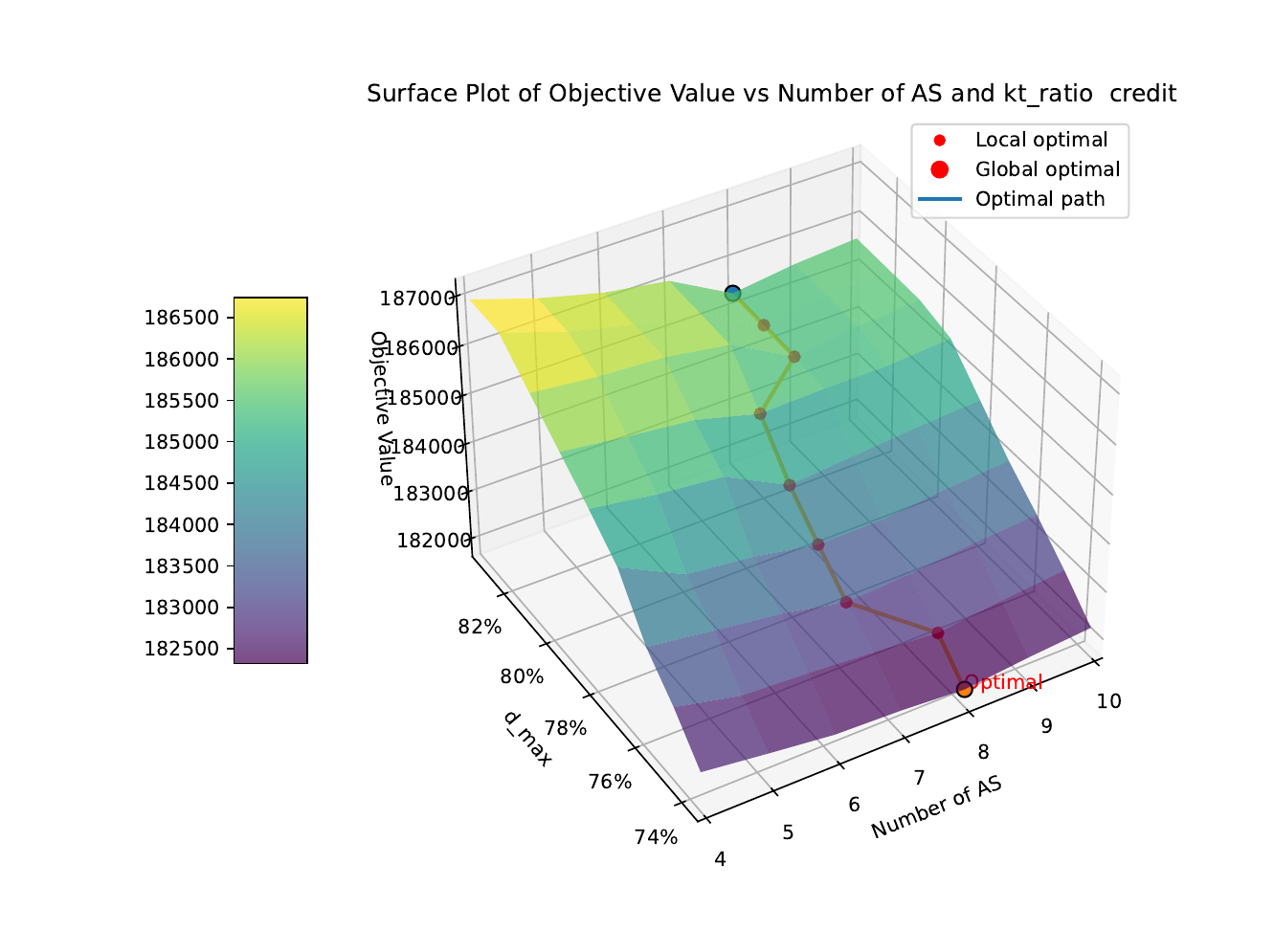}
    \caption{The solution surface over $d_{max}$ and number of DRAS}
    \label{fig:even incentive}
\end{figure}

We first present the result without considering of the impact of credit price on travelers' behavior. To illustrate the optimization process and the structure of the solution space, we visualize the objective values over all tested combinations of the maximum allowable driving ratio ($d_{max}$) and the DRAS fleet size. Figure~\ref{fig:even incentive} presents the resulting 3D surface, where each point represents a configuration of system parameters and the corresponding objective outcome. The red markers denote locally optimal configurations for each TCS credit ratio, while the larger red marker highlights the global optimum within the search domain.

As shown in \autoref{fig:even incentive}, objective value improves significantly as $d_{max}$ decreases from $83\%$ to $74\%$. 
In the absence of the TCS policy intervention, approximately $81\%$ of travelers choose a private car for travel. As $d_{max}$ ratio drops below this rate, the credit constraint becomes binding, discouraging car users from driving and prompting mode shifts to buses or DRASs. 

Interestingly, the required DRAS fleet size to achieve local optimality shows a non-monotonic pattern. When maximum allowable driving rate is mild ($d_{max} \approx 80\%$), the system requires a relatively large DRAS fleet (e.g., $8$ vehicles) to compensate for high private car usage and its associated externalities. Once TCS is activated and the credit budget is bounded (i.e., $d_{max} < 80\%$), a portion of car users switch to public transit. This consequence improves system-level efficiency by curbing excessive congestion and emission, and the displaced demand is expected absorbed by shared transport without triggering excessive queuing time and delay. As TCS becomes stricter (i.e., as $d_{max}$ get close to $ 76\%$), a large number of travelers are shifted from private cars, which beyond the system can accommodate, leading to longer queues. In response, the system may compensate by deploying some DRAS fleet to absorb some of the demand and mitigate these negative externalities. This dynamic illustrates how TCS and DRAS investment can act as substitutes at first, and complements later, depending on the strength of the constraint.

To further illustrate this mechanism, we adjust the credit redemption scheme. Under the modified scheme, bus trips will earn half of the credit price multiplied by the initially allocated amount, which can be expressed as \(0.5 \times k \times \text{credit price}\). In contrast, DRAS trips will earn one full credit price multiplied by the initial allocated amount, or \(1 \times k \times \text{credit price}\). This adjustment increases the relative attractiveness of DRAS for travelers displaced from private cars when bus capacity is saturated, showing more clearly how DRAS absorbs the residual demand that cannot be accommodated by buses under stricter TCS settings.

\begin{figure}[htbp]
    \centering
    \includegraphics[width=0.95\linewidth]{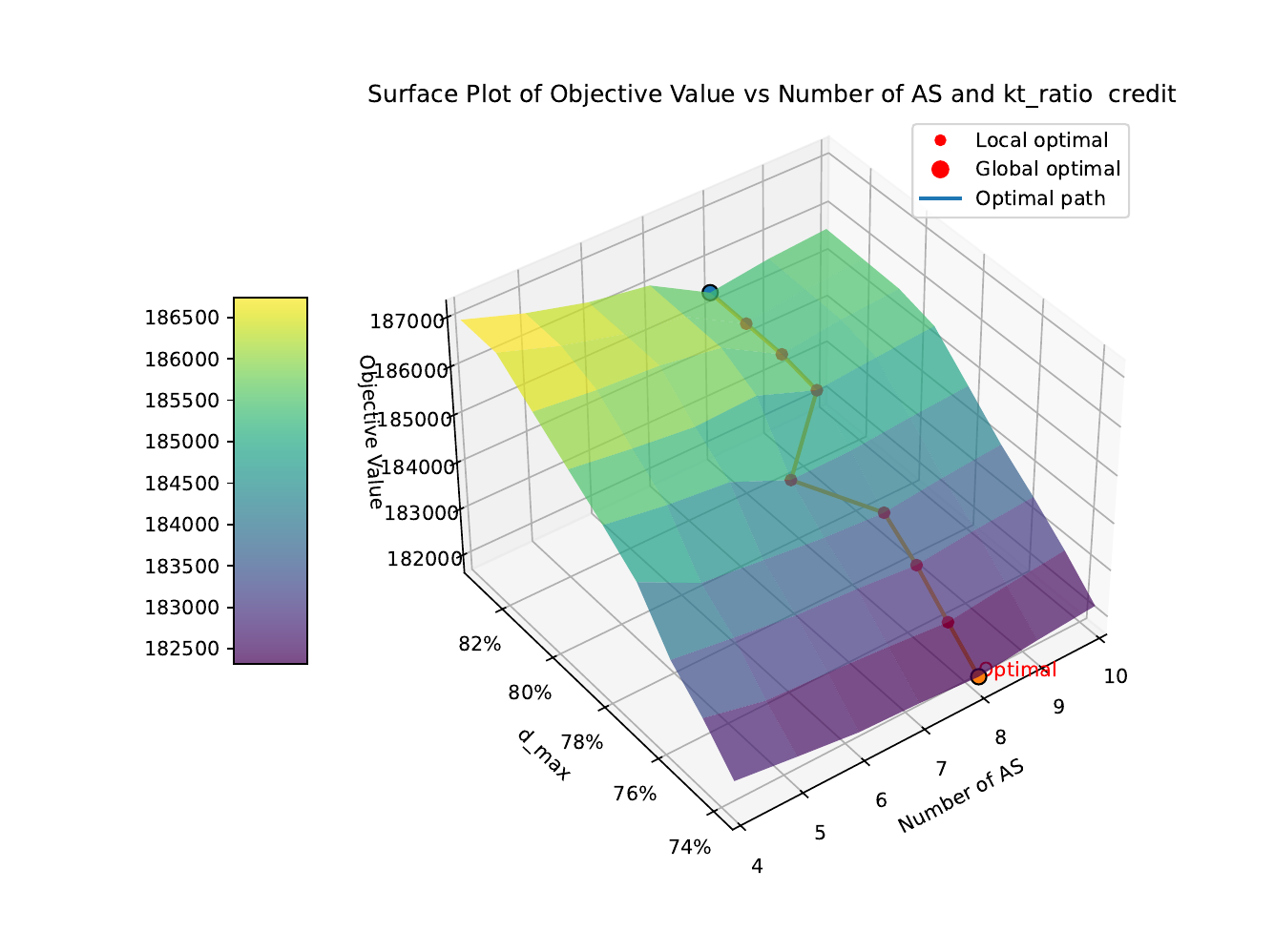}
    \caption{The solution surface over $d_{max}$ and number of DRAS with higher monetary compensation to DRAS riders}
    \label{fig:placeholder2}
\end{figure}
As \autoref{fig:placeholder2} shows, the required DRAS fleet size to achieve local optimality exhibits a non-monotonic pattern, but the required number of DRAS for optimal is relatively higher, demonstrating that TCS and DRAS act as substitutes at first, and complements later.

\subsubsection*{Results considering credit charge}
\begin{figure}
    \centering
    \includegraphics[width=1\linewidth]{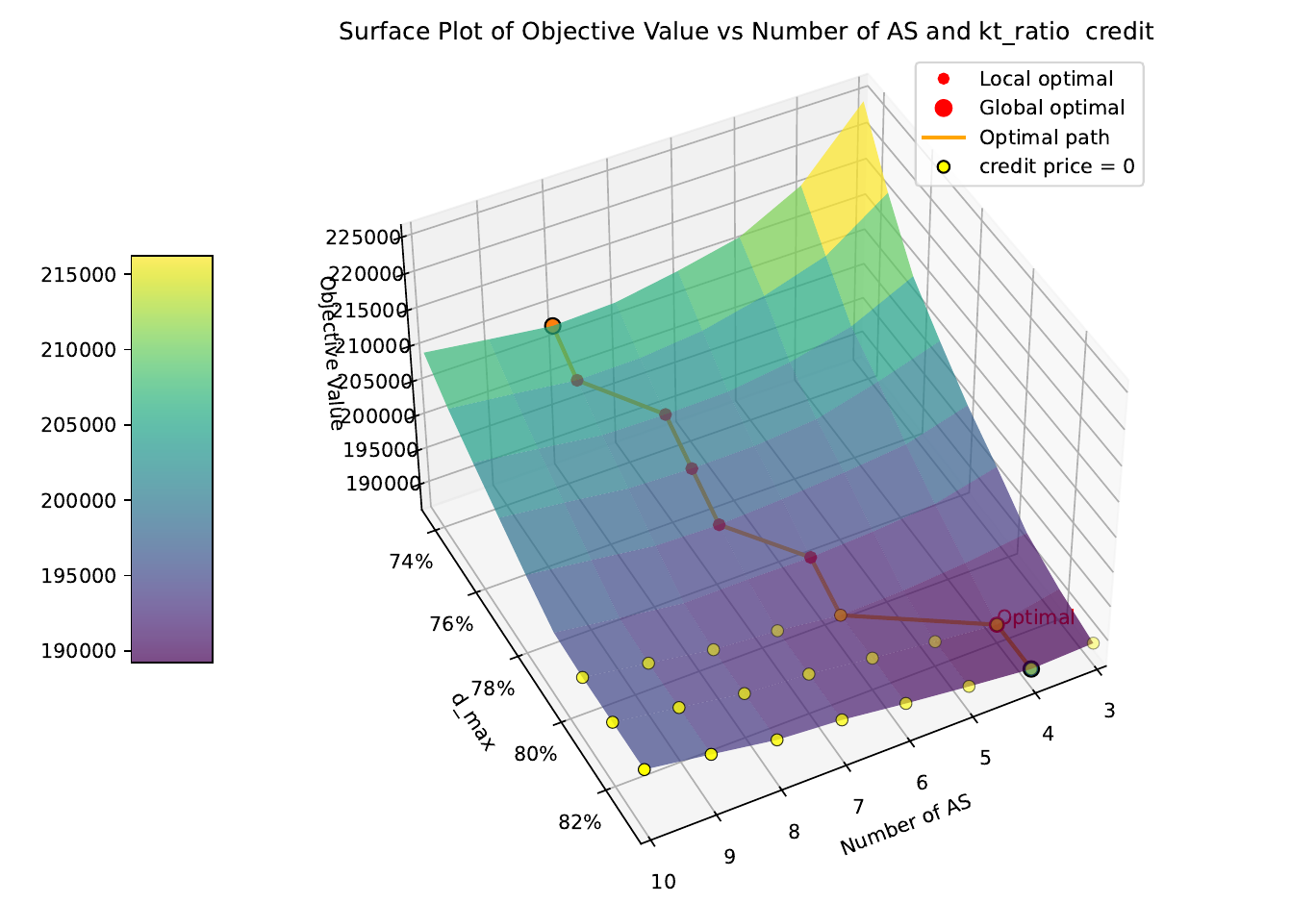}
    \caption{The solution surface over $d_{max}$ and number of DRAS considering credit price impact}
    \label{fig:same k long}
\end{figure}

After incorporating the credit price term, the entire result surface exhibits a notable inversion compared to the results without price consideration (see \autoref{fig:same k long} as a reference). Previously, the lowest (i.e., 74\%) driving allowable rate corresponded to a higher objective value, suggesting that restricting private car use led to better overall outcomes. However, once travelers' concern to credit price is introduced, the situation reverses. Although reducing the driving rate can still improve network performance, travelers now tend to avoid expensive options. As a result, the optimization shifts toward solutions that minimize individual spending, even at the cost of slightly lower system-wide efficiency.

This adjustment causes the optimization model to favor scenarios with fewer or no credit purchases (i.e., credit price = 0 points). Consequently, the global optimum now lies in a region that balances reduced travel costs with moderate network performance, rather than purely maximizing the objective value of the system performance as before. We pick the lowest points under the effects of both TCS and DRAS implemention (i.e., DRAS = 6, and TCS = 79\%) as the optimal result. (see \autoref{tab:optimal_summary} as a reference)

In summary, introducing price consideration flips the optimization landscape. The objective surface becomes distorted by travelers’ cost aversion, leading to more realistic and economically viable outcomes. 

\begin{table}[htbp]
\centering
\caption{System Performance and Mode Shares under Combined DRAS Implementation and TCS Scenario}
\small    
\begin{tabular}{@{}lr@{}}
\toprule
\textbf{Metric} & \textbf{Value} \\
\midrule
Number of DRAS                       & 6 \\
Total Travel Time (h)           & 16,934.25 \\
Total Distance (km)                  & 391,723.4 \\
DRAS Deployment Cost (\euro{}/vehicle) & 6.0M \\
Objective Value (\euro{})            & 193220 \\
Car Mode Share                & 79\% \\
Credit Price                         & 0.1236\,\euro{} \\
Total Credit Cost for Driving        & 1.60\,\euro{} \\
\midrule
OD 0--2 Mode Share (Car / Bus / DRAS) & 79.8\% / 11.1\% / 9.6\% \\
OD 1--2 Mode Share (Car / Bus / DRAS) & 77.9\% / 14.2\% / 8.0\% \\
OD 0--1 Mode Share (Car / Bus / DRAS) & 79.2\% / 11.0\% / 9.8\% \\
\bottomrule
\end{tabular}
\normalsize
\label{tab:optimal_summary}
\end{table}


\subsection{Result comparision among different combination of policies}
To deepen the analysis, we pick the optimal point that under the effects of both TCS and DRAS implemention (i.e., DRAS = 6, and TCS = 79\%) as the study point, and further evaluate four distinct policy scenarios: (i) a baseline with no intervention, (ii) implementation of TCS alone, (iii) deployment of DRAS without TCS, and (iv) a combined strategy integrating both TCS and DRAS. 


\begin{table}[htbp]
\centering
\caption{System Performance under Different Policy Scenarios}
\resizebox{\textwidth}{!}{%
\begin{tabular}{llrrrrrrc}
\toprule
\textbf{DRAS} & \textbf{TCS} &
\makecell{\#\\DRAS} &
\makecell{Travel Time\\(h)} &
\makecell{Distance\\(km)} &
\makecell{DRAS Cost\\(\euro{})} &
\makecell{Credit\\Price\\(\euro{})} &
\makecell{Charge per\\Trip (\euro{}/trip)} &
\makecell{Car\\Share} \\
\midrule
No  & No  & 0  & 18,826.61 & 421,836.04 & 0.0M (0 per day)    & 0.00   & 0.00   & 85\%\\
Yes & No  & 6  & 17,736.65 & 403,632.17 & 6.0M (1643 per day) & 0.00   & 0.00  & 81\% \\
No  & Yes & 0  & 17,357.53 & 403,073.96 & 0.0M (0 per day)    & 0.62   & 8.11   & 79\% \\
Yes & Yes & \textbf{6} &
\textbf{16,934.25} &
\textbf{391,723.4} &
\textbf{6.0M (1643 per day)} &
\textbf{0.1236} &
\textbf{1.60} &
\textbf{79\%} \\
\bottomrule
\label{withpricetable}
\end{tabular}%
}

\begin{tablenotes}
\small
\item \textit{Note:} The last row corresponds to the scenario where both DRAS and TCS are implemented. This setting yields the lowest system cost among all scenarios considered, indicating the most cost-effective system performance.
\end{tablenotes}
\end{table}

As shown in the \autoref{withpricetable}, the combined implementation of TCS and DRAS yields the most balanced and effective outcome in the selected scenarios. Deploying DRASs alone improves system efficiency through supply-side enhancement, reducing both total travel time and mileage. However, over 81\% of travelers still choose private cars for traveling, indicating that additional demand-responsive service may not induce substantial modal shifts in the absence of regulatory constraints. On the other hand, the TCS-only scenario achieves a reduction in car mode share (around 6\%) by directly discouraging private car use. Yet, its effect on total system performance is limited. This is likely because the displaced car users are fostered to already constrained public and shared modes, potentially leading to service bottlenecks and extended waiting times. The integrated TCS + DRAS strategy addresses both demand and supply dimensions simultaneously. It maintains the behavioral impact of TCS while alleviating capacity pressure through flexible DRAS deployment. As a result, this scenario delivers the lowest system cost, along with a 7.3\% reduction in travel time and a 7.0\% drop in total vehicle kilometers compared to the baseline. 

\subsection{Summary and discussion} 
Our numerical analysis demonstrates that the proposed models are capable of capturing complex and non-monotonic interactions between demand-side regulation (through TCS), supply-side operational mechanism (via both DRAS deployment and bus operations), and resulting user equilibrium behaviors.

Under the single OD scenario, we reveal that tightening TCS (i.e., lowering the $d_{max}$) ratio shifts the credit market from slack to binding, raising credit prices, discouraging private car use, and reducing in-vehicle times through congestion relief. However, the displaced demand are not sufficient served by the bus or DRAS system create boarding waiting time, but the overall the congestion benefits generally outweigh the queuing costs in our corridor setting, leading to net reductions in total travel time. Furthermore, expanding the DRAS fleet primarily reduces queuing delays rather than altering in-vehicle speeds, with clear diminishing returns once initial capacity shortages are addressed. 
In the multi-OD bilevel framework, endogenizing credit prices in upper level objective function reshapes the optimization landscape; The objective surface is distorted by travelers’ cost concern, resulting in a preference for scenarios with fewer or no credit purchases. In our policy comparison across four scenarios, no policy, TCS only, DRAS only, and TCS+DRAS, the joint implementation (TCS+DRAS) delivers the lowest system cost.



\section{Conclusion}
In this paper, we develop a dynamic multimodal equilibrium framework that considers operational constraints of public or shared transport services under TCS regulation. Unlike prior studies that often assume unlimited transit capacity or exogenous transit supply, our model explicitly incorporates key service features, including vehicle capacity, fleet size, and departure frequency, while capturing congestion effects through a multimodal MFD. To describe station-level crowding and boarding delays under time-varying demand, we incorporate a point queue model that dynamically tracks passenger accumulation and vehicle dispatching at each station, considering key service features. The formulations capture how travelers respond to credit pricing and service availability within a stochastic user equilibrium framework, and how those responses, in turn, influence system-level efficiency and modal choices.

From the supply side, we introduce a DRAS as a flexible and adaptive alternative. We further develop a bi-level optimization framework (reduced to MPEC) that jointly determines policy parameters (credit allocations and charging rules) and operational decisions (DRAS fleet sizing). The upper-level problem reflects a central planner's objective to minimize a weighted combination of system-wide travel time, travel distance, investment costs, and credit price, while the lower-level captures user equilibrium under endogenous network and service conditions. This structure enables policy evaluation and coordinated design of pricing and mobility services in a unified, behaviorally responsive framework.

Our numerical results on a corridor segment of the A10 highway near Paris reveal several key insights. First, the proposed integrated optimization–equilibrium framework is capable of capturing complex mechanisms among different modules. Second, the implementation of TCS alone can significantly reduce private car usage by raising its generalized cost; however, its effectiveness depends critically on the availability of shared alternatives due to the waiting effect. This insight reveals that the TCS is not always self-sufficient; without matching capacity adjustments, it may create congestion in shared alternatives and deteriorate the system efficiency gains from demand regulation. Third, introducing DRAS as a flexible supply alternative alleviates the pressure on both traffic congestion and fixed-schedule transit and enables a more elastic response to peak demand. We observe that DRAS deployment not only reduces total travel time and emissions, but also smoothens price fluctuations in the credit market by absorbing displaced demand.

Most notably, our results highlight a non-monotonic interaction between TCS and DRAS deployment. In early phases of credit restriction, the behavioral filtering induced by TCS discourages car usage, improving overall traffic conditions without requiring additional investment. As credit tightness increases further, however, the acceptance of TCS could be bad, and shared modes may become saturated, and additional DRAS supply becomes necessary to preserve service quality and maintain accessibility. This interplay illustrates a fundamental complementarity between policy levers (TCS) and operational flexibility (DRAS), with implications for system efficiency.

From a methodological standpoint, the proposed model uncovers unintended bottlenecks, substitution effects between policies, and the inflection points at which additional investment becomes critical. For practitioners, our findings underscore the importance of coordinated design: demand policy must be supported by adequate supply-side interventions, and infrastructure investments must anticipate the behavioral responses they aim to elicit. Future extensions could incorporate user heterogeneity, dynamic credit trading behavior, or adaptive learning models to further explore temporal effects and long-term policy impacts. 

\begin{table}[htbp]
\centering
\caption{System Performance under Different Policy Scenarios}
\resizebox{\textwidth}{!}{%
\begin{tabular}{llrrrrrrcc}
\toprule
\textbf{DRAS} & \textbf{TCS} &
\makecell{\#\\DRAS} &
\makecell{Travel Time\\(h)} &
\makecell{$\Delta$TT\\(\%)} &
\makecell{Distance\\(km)} &
\makecell{$\Delta$Dist\\(\%)} &
\makecell{Charge per\\Trip (\euro{}/trip)} &
\makecell{Car\\Share} &
\makecell{DRAS\\Share} \\
\midrule
No  & No  & 0  & $1.88{\times}10^{4}$ & 0\%    & $4.22{\times}10^{5}$ & 0\%    & 0.00  & 85\% & 0\% \\
Yes & No  & 6  & $1.77{\times}10^{4}$ & -5.8\% & $4.04{\times}10^{5}$ & -4.3\% & 0.00  & 81\% & 2.9\%\\
No  & Yes & 0  & $1.74{\times}10^{4}$ & -7.8\% & $4.03{\times}10^{5}$ & -4.4\% & 8.11  & 79\% & 0\%\\
Yes & Yes & \textbf{6} &
\textbf{$1.69{\times}10^{4}$} &
\textbf{-10.0\%} &
\textbf{$3.92{\times}10^{5}$} &
\textbf{-7.1\%} &
\textbf{1.60} &
\textbf{79\%} &
\textbf{3.7\%} \\
\bottomrule
\label{withpricetable}
\end{tabular}%
}

\begin{tablenotes}
\small
\item \textit{Note:} Travel time and distance are reported using scientific notation. The No–No scenario is used as the baseline.
\end{tablenotes}
\end{table}

\section*{CRediT authorship contribution statement}
\textbf{Zenghao Hou:} Conceptualization of this study, Investigation, Methodology, Coding, Formal analysis, Visualization, Writing – original draft, Writing – review and editing. 
\textbf{Ludovic Leclercq:} Conceptualization of this study, Methodology, Supervision, Writing – review and editing, Funding support.

\section*{Acknowledgment}
This work was performed within the MOBAUTO project, supported by the Future Investment Program initiated by the French government and operated by Bpifrance, the French public investment bank within France 2030.
\bibliography{cas-sc-template.bib}

@article{wardman2004public,
  title={Public transport values of time},
  author={Wardman, Mark},
  journal={Transport policy},
  volume={11},
  number={4},
  pages={363--377},
  year={2004},
  publisher={Elsevier}
}

@article{fosgerau2007danish,
  title={The danish value of time study},
  author={Fosgerau, Mogens and Hjorth, Katrine and Lyk-Jensen, St{\'e}phanie Vincent},
  year={2007},
  publisher={The Danish Transport Research Institute}
}

@article{xiao2013managing,
  title={Managing bottleneck congestion with tradable credits},
  author={Xiao, Feng and Qian, Zhen Sean and Zhang, H Michael},
  journal={Transportation Research Part B: Methodological},
  volume={56},
  pages={1--14},
  year={2013},
  publisher={Elsevier}
}

@techreport{cook2025short,
  title={The Short-Run Effects of Congestion Pricing in New York City},
  author={Cook, Cody and Kreidieh, Aboudy and Vasserman, Shoshana and Allcott, Hunt and Arora, Neha and van Sambeek, Freek and Tomkins, Andrew and Turkel, Eray},
  year={2025},
  institution={National Bureau of Economic Research}
}

@article{bao2019regulating,
  title={Regulating dynamic congestion externalities with tradable credit schemes: Does a unique equilibrium exist?},
  author={Bao, Yue and Verhoef, Erik T and Koster, Paul},
  journal={Transportation Research Part B: Methodological},
  volume={127},
  pages={225--236},
  year={2019},
  publisher={Elsevier}
}

@article{vickrey1969congestion,
  title={Congestion theory and transport investment},
  author={Vickrey, William S},
  journal={The American economic review},
  volume={59},
  number={2},
  pages={251--260},
  year={1969},
  publisher={JSTOR}
}

@article{yang2011managing,
  title={Managing network mobility with tradable credits},
  author={Yang, Hai and Wang, Xuesong},
  journal={Transportation Research Part B: Methodological},
  volume={45},
  number={3},
  pages={580--594},
  year={2011},
  publisher={Elsevier}
}

@article{tian2013tradable,
  title={Tradable credit schemes for managing bottleneck congestion and modal split with heterogeneous users},
  author={Tian, Li-Jun and Yang, Hai and Huang, Hai-Jun},
  journal={Transportation Research Part E: Logistics and Transportation Review},
  volume={54},
  pages={1--13},
  year={2013},
  publisher={Elsevier}
}

@article{zhang2013modelling,
  title={Modelling network flow with and without link interactions: the cases of point queue, spatial queue and cell transmission model},
  author={Zhang, HM and Nie, Yu and Qian, Zhen},
  journal={Transportmetrica B: Transport Dynamics},
  volume={1},
  number={1},
  pages={33--51},
  year={2013},
  publisher={Taylor \& Francis}
}

@article{wu2012design,
  title={Design of more equitable congestion pricing and tradable credit schemes for multimodal transportation networks},
  author={Wu, Di and Yin, Yafeng and Lawphongpanich, Siriphong and Yang, Hai},
  journal={Transportation Research Part B: Methodological},
  volume={46},
  number={9},
  pages={1273--1287},
  year={2012},
  publisher={Elsevier}
}

@article{nagurney2000marketable,
  title={Marketable pollution permits in oligopolistic markets with transaction costs},
  author={Nagurney, Anna and Dhanda, Kanwalroop Kathy},
  journal={Operations Research},
  volume={48},
  number={3},
  pages={424--435},
  year={2000},
  publisher={INFORMS}
}

@article{tang2021cost,
  title={The cost of traffic: evidence from the London congestion charge},
  author={Tang, Cheng Keat},
  journal={Journal of Urban Economics},
  volume={121},
  pages={103302},
  year={2021},
  publisher={Elsevier}
}

@article{daganzo2001simple,
  title={A simple traffic analysis procedure},
  author={Daganzo, Carlos F},
  journal={Networks and Spatial Economics},
  volume={1},
  pages={77--101},
  year={2001},
  publisher={Springer}
}

@article{verhoef1997tradeable,
  title={Tradeable permits: their potential in the regulation of road transport externalities},
  author={Verhoef, Erik and Nijkamp, Peter and Rietveld, Piet},
  journal={Environment and Planning B: planning and design},
  volume={24},
  number={4},
  pages={527--548},
  year={1997},
  publisher={SAGE Publications Sage UK: London, England}
}

@article{balzer2023dynamic,
  title={Dynamic tradable credit scheme for multimodal urban networks},
  author={Balzer, Louis and Ameli, Mostafa and Leclercq, Ludovic and Lebacque, Jean-Patrick},
  journal={Transportation Research Part C: Emerging Technologies},
  volume={149},
  pages={104061},
  year={2023},
  publisher={Elsevier}
}

@article{balzer2022modal,
  title={Modal equilibrium of a tradable credit scheme with a trip-based MFD and logit-based decision-making},
  author={Balzer, Louis and Leclercq, Ludovic},
  journal={Transportation Research Part C: Emerging Technologies},
  volume={139},
  pages={103642},
  year={2022},
  publisher={Elsevier}
}

@article{nie2013managing,
  title={Managing rush hour travel choices with tradable credit scheme},
  author={Nie, Yu Marco and Yin, Yafeng},
  journal={Transportation Research Part B: Methodological},
  volume={50},
  pages={1--19},
  year={2013},
  publisher={Elsevier}
}

@article{frey1996old,
  title={The old lady visits your backyard: A tale of morals and markets},
  author={Frey, Bruno S and Oberholzer-Gee, Felix and Eichenberger, Reiner},
  journal={Journal of political economy},
  volume={104},
  number={6},
  pages={1297--1313},
  year={1996},
  publisher={The University of Chicago Press}
}

@article{arnott1994welfare,
  title={The welfare effects of congestion tolls with heterogeneous commuters},
  author={Arnott, Richard and De Palma, Andr{\'e} and Lindsey, Robin},
  journal={Journal of Transport Economics and Policy},
  pages={139--161},
  year={1994},
  publisher={JSTOR}
}

@article{chen2023market,
  title={Market design for tradable mobility credits},
  author={Chen, Siyu and Seshadri, Ravi and Azevedo, Carlos Lima and Akkinepally, Arun P and Liu, Renming and Araldo, Andrea and Jiang, Yu and Ben-Akiva, Moshe E},
  journal={Transportation Research Part C: Emerging Technologies},
  volume={151},
  pages={104121},
  year={2023},
  publisher={Elsevier}
}

@article{gu2018congestion,
  title={Congestion pricing practices and public acceptance: A review of evidence},
  author={Gu, Ziyuan and Liu, Zhiyuan and Cheng, Qixiu and Saberi, Meead},
  journal={Case Studies on Transport Policy},
  volume={6},
  number={1},
  pages={94--101},
  year={2018},
  publisher={Elsevier}
}

@article{leclercq2017dynamic,
  title={Dynamic macroscopic simulation of on-street parking search: A trip-based approach},
  author={Leclercq, Ludovic and S{\'e}n{\'e}cat, Alm{\'e}ria and Mariotte, Guilhem},
  journal={Transportation Research Part B: Methodological},
  volume={101},
  pages={268--282},
  year={2017},
  publisher={Elsevier}
}

@article{lamotte2018morning,
  title={The morning commute in urban areas with heterogeneous trip lengths},
  author={Lamotte, Rapha{\"e}l and Geroliminis, Nikolas},
  journal={Transportation Research Part B: Methodological},
  volume={117},
  pages={794--810},
  year={2018},
  publisher={Elsevier}
}

@article{fan2016waiting,
  title={Waiting time perceptions at transit stops and stations: Effects of basic amenities, gender, and security},
  author={Fan, Yingling and Guthrie, Andrew and Levinson, David},
  journal={Transportation Research Part A: Policy and Practice},
  volume={88},
  pages={251--264},
  year={2016},
  publisher={Elsevier}
}

@article{ji2019waiting,
  title={Waiting time perceptions at bus and metro stations in Nanjing, China: the importance of station amenities, trip contexts, and passenger characteristics},
  author={Ji, Yanjie and Gao, Liangpeng and Fan, Yingling and Zhang, Chu and Zhang, Ruochen},
  journal={Transportation Letters},
  volume={11},
  number={9},
  pages={479--485},
  year={2019},
  publisher={Taylor \& Francis}
}

@misc{facchinei2003finite,
  title={Finite-dimensional variational inequalities and complementarity problems},
  author={Facchinei, F},
  year={2003},
  publisher={Springer}
}

@article{jin2020generalized,
  title={Generalized bathtub model of network trip flows},
  author={Jin, Wen-Long},
  journal={Transportation Research Part B: Methodological},
  volume={136},
  pages={138--157},
  year={2020},
  publisher={Elsevier}
}

@article{dirkse1995path,
  title={The path solver: a nommonotone stabilization scheme for mixed complementarity problems},
  author={Dirkse, Steven P and Ferris, Michael C},
  journal={Optimization methods and software},
  volume={5},
  number={2},
  pages={123--156},
  year={1995},
  publisher={Taylor \& Francis}
}

@article{mariotte2017macroscopic,
  title={Macroscopic urban dynamics: Analytical and numerical comparisons of existing models},
  author={Mariotte, Guilhem and Leclercq, Ludovic and Laval, Jorge A},
  journal={Transportation Research Part B: Methodological},
  volume={101},
  pages={245--267},
  year={2017},
  publisher={Elsevier}
}

\end{document}